\documentclass{aa} 
\usepackage{xcolor}
\usepackage{support-caption}
\usepackage{subcaption}
\usepackage{graphicx}

\newcommand{\civalone}{\textrm{C}\textsc{iv}}

\newcommand{\ha}{\ifmmode {\rm H}\alpha \else H$\alpha$\fi}
\newcommand{\hb}{\ifmmode {\rm H}\beta \else H$\beta$\fi}
\newcommand{\lya}{\ifmmode {\rm Ly}\alpha \else Ly$\alpha$\fi}
\newcommand{\pg}{\ifmmode {\rm P}\gamma \else Pa$\gamma$\fi}
\newcommand{\lyb}{\ifmmode {\rm Ly}\beta \else Ly$\beta$\fi}
\newcommand{\lyg}{\ifmmode {\rm Ly}\gamma \else Ly$\gamma$\fi}

\newcommand{\flyc}{\ifmmode \mathrm{f}_\mathrm{esc}\mathrm{(LyC)} \else $\mathrm{f}_\mathrm{esc}\mathrm{(LyC)}$\fi}

\def\ergs{\ifmmode \mathrm{erg\hspace{1mm}s}^{-1} \else erg s$^{-1}$\fi}

\def\micron{\ifmmode \mu\mathrm{m} \else $\mu$m\fi}
\def\msun{\ifmmode \mathrm{M}_{\odot} \else M$_{\odot}$\fi}
\def\msunyr{\ifmmode \mathrm{M}_{\odot} \hspace{1mm}{\rm yr}^{-1} \else $\mathrm{M}_{\odot}$ yr$^{-1}$\fi}
\def\zsun{\ifmmode Z_{\odot} \else Z$_{\odot}$\fi}
\def\lsun{\ifmmode L_{\odot} \else L$_{\odot}$\fi}
\def\mstar{\ifmmode \mathrm{M}_{\star} \else M$_{\star}$\fi}

\usepackage{txfonts}
\newcommand{\orcid}[1]{\href{https://orcid.org/#1}{\includegraphics[width=10pt]{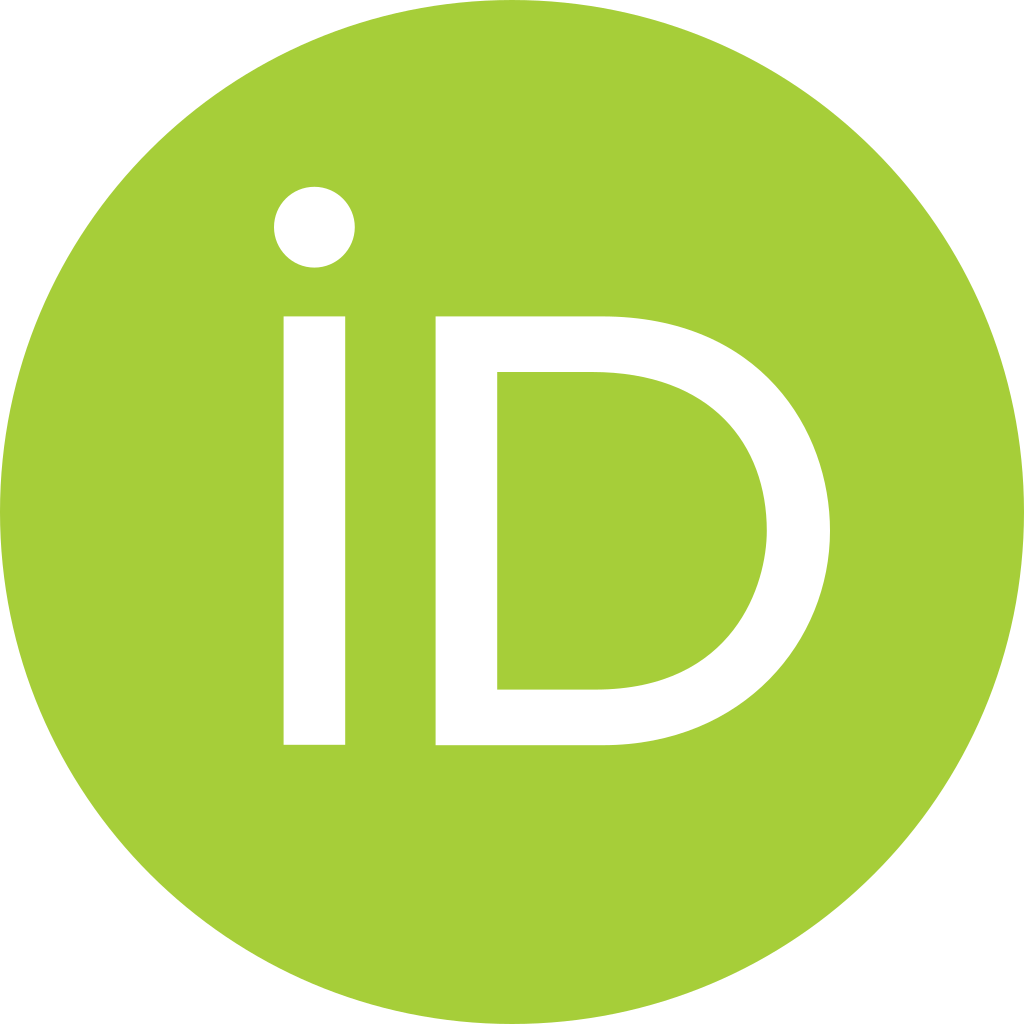}}}
\usepackage{hyperref}
\hypersetup{colorlinks, 
  linkcolor = [rgb]{1.,0.2,0.2},
  urlcolor = [rgb]{0.7,0.2,0.2},
  citecolor = [rgb]{0.1,0.4,1.},
  anchorcolor = [rgb]{0.7,0.2,0.2}
}
\begin{document} 

 \title{Insights into the reionization epoch from cosmic-noon-\textrm{C}\textsc{iv} emitters in the VANDELS survey
}

 \subtitle{}
 \author{S. Mascia \orcid{0000-0002-9572-7813}
 \inst{1,2}
 \and L. Pentericci \orcid{0000-0001-8940-6768}
 \inst{1}
  \and A. Saxena \orcid{0000-0001-5333-9970}
 \inst{3,4}
 \and 
 D. Belfiori 
 \inst{1}
 \and
 A. Calabrò \orcid{0000-0003-2536-1614}
 \inst{1}
 \and M. Castellano \orcid{0000-0001-9875-8263}
 \inst{1}
 \and 
 A. Saldana-Lopez \orcid{0000-0001-8419-3062}
 \inst{7}
\and
 M. Talia \orcid{0000-0003-4352-2063}
 \inst{5}
 \and
  R. Amorín \orcid{0000-0001-5758-1000}
 \inst{8,9}
 \and 
 F. Cullen \orcid{0000-0002-3736-476X}
 \inst{10}
 \and
 B. Garilli \orcid{0000-0001-7455-8750}
 \inst{11}
 L. Guaita \orcid{0000-0002-4902-0075}
 \inst{12}
 \and
 M. Llerena \orcid{0000-0003-1354-4296}
\inst{9}
\and
R. J. McLure
\inst{10}
\and
 M. Moresco \orcid{0000-0002-7616-7136}
 \inst{6} 
 \and
 P. Santini \orcid{0000-0002-9334-8705}
 \inst{1}
 \and
 D. Schaerer \orcid{0000-0001-7144-7182}
 \inst{7}
 }
 
 \institute{\textit{INAF – Osservatorio Astronomico di Roma, via Frascati 33, 00078, Monteporzio Catone, Italy}\\
  \email{sara.mascia@inaf.it}
  \and 
  \textit{Dipartimento di Fisica, Università di Roma Tor Vergata,
Via della Ricerca Scientifica, 1, 00133, Roma, Italy}
\and 
\textit{Sub-department of Astrophysics, University of Oxford, Keble Road, Oxford OX1 3RH, United Kingdom}
\and
\textit{Department of Physics and Astronomy, University College London, Gower Street, London WC1E 6BT, UK}
\and
\textit{University of Bologna - Department of Physics and Astronomy “Augusto Righi” (DIFA), Via Gobetti 93/2, I-40129, Bologna, Italy}
\and
\textit{INAF - Osservatorio Astronomico di Bologna, via P. Gobetti 93/3, I-40129, Bologna, Italy}
\and 
\textit{Department of Astronomy, University of Geneva, 51 Chemin Pegasi, 1290 Versoix, Switzerland
 ASL acknowledge support from Swiss National Science Foundation} 
 \and
 \textit{Instituto de Investigación Multidisciplinar en Ciencia y Tecnología, Universidad de La Serena, Raúl Bitrán 1305, La Serena, Chile}
 \and 
\textit{Departamento de Física y Astronomía, Universidad de La Serena, Av. Juan Cisternas 1200 Norte, La Serena, Chile}
  \and
  \textit{SUPAScottish Universities Physics Alliance, Institute for Astronomy, University of Edinburgh, Royal Observatory, Edinburgh EH9 3HJ}
  \and
  \textit{INAF-IASF Milano, Via Alfonso Corti 12, I-20133 Milano,
Italy}
\and
\textit{Departamento de Ciencias Fisicas, Facultad de Ciencias Exactas, Universidad Andres Bello, Fernandez Concha 700, Las Condes, Santiago, Chile
}
}
 \date{Accepted XXX. Received YYY; in original form ZZZ}

\abstract{Recently, intense emission from nebular \textrm{C}\textsc{iii}] and \textrm{C}\textsc{iv} emission lines have been observed in galaxies in the epoch of reionization ($z>6$) and have been proposed as the prime way of measuring their redshift and studying their stellar populations. These galaxies might represent the best examples of cosmic reionizers, as suggested by recent low-z observations of Lyman Continuum emitting galaxies, but it is hard to directly study the production and escape of ionizing photons at such high redshifts. The ESO spectroscopic public survey VANDELS offers the unique opportunity to find rare examples of such galaxies at cosmic noon ($z\sim 3$), thanks to the ultra deep observations available. We have selected a sample of 39 galaxies showing \textrm{C}\textsc{iv} emission, whose origin (after a careful comparison to photoionization models) can be ascribed to star formation and not to AGN. By using a multi-wavelength approach, we determine their physical properties including metallicity and ionization parameter and compare them to the properties of the parent population to understand what are the ingredients that could characterize the analogs of the cosmic reionizers. We find that \textrm{C}\textsc{iv} emitters are galaxies with high photons production efficiency and there are strong indications that they might have also large escape fraction: given the visibility of \textrm{C}\textsc{iv} in the epoch of reionization this could become the best tool to pinpoint the cosmic reioinzers.}
 


 \keywords{galaxies: evolution - galaxies: high-redshift – dark ages, reionization, first stars – early Universe}

 \maketitle
\section{Introduction}
The rest-frame UV and optical emission lines that emerge from star forming galaxies are an important indicator of the nature and strength of the ionizing radiation and the conditions in the Inter-Stellar Medium (ISM) since they can be used to infer gas-phase abundances and ionization parameters.
Particularly the rest-UV range is currently fundamental to study the early galaxies in the epoch of reionization (EoR), as it is accessible with ground-based telescopes and now with the \textit{JWST} \citep{Robertson_2021}. In the last few years, prominent high-ionization nebular emission lines such as \textrm{O}\textsc{iii}] $\lambda\lambda$ 1661,1666, \textrm{C}\textsc{iv} $\lambda\lambda$ 1548,1551 (hereafter simply \textrm{C}\textsc{iv} $\lambda$ 1550), \textrm{He}\textsc{ii} $\lambda$ 1640 and \textrm{C}\textsc{iii}] $\lambda$ 1907 + \textrm{C}\textsc{iii}] $\lambda$ 1909 (hereafter simply \textrm{C}\textsc{iii}] $\lambda$ 1910) have been observed in a few $z \simeq 6-7$ galaxies \citep{Stark_2015,Mainali_2018}, indicating that these systems are characterised by extreme radiation fields. Whether such properties are really common or not at these early epochs is still a matter of debate \citep[e.g. see][]{Castellano_2022}, but they might become the best probes to detect and characterise the most distant sources, since at $z>7$ the Ly$\alpha$ emission line, usually the strongest line in young star forming galaxies, starts to be suppressed by the increasingly neutral Intergalactic Medium (IGM) as we approach the reionization epoch \citep{Pentericci_2018b,jung2020,Ouchi_2020,bolan2021}.

The nebular \textrm{C}\textsc{iv} emission line, requiring comparably high ionisation energies as the \textrm{He}\textsc{ii} line (E $>$ 47.9 eV and $>$ 49.9 eV respectively), has recently attracted attention since it has been shown that its presence might be strongly linked to the escape of Lyman continuum photons from galaxies. Indeed \cite{Schaerer_2022} recently detected \textrm{C}\textsc{iv} in several confirmed Lyman continuum (LyC) leakers at $z < 0.7$, finding \textrm{C}\textsc{iv} emission in all LyC emitters with escape fractions $f_{esc}>0.1$ and suggesting that such strong leakers have \textrm{C}\textsc{iv}/\textrm{C}\textsc{iii} ratios above $\sim$ 0.75. This makes the \textrm{C}\textsc{iv} emission line a very promising indirect tracer of LyC $f_{esc}$.

 Other observations also support this scenario: one of the few solid Lyman continuum leakers at high redshift, Ion2 at $z = 3.2$, shows the presence of nebular emission in both \textrm{C}\textsc{iv} and \textrm{C}\textsc{iii}], with a rather high ratio \textrm{C}\textsc{iv}/\textrm{C}\textsc{iii}] from deep X-Shooter spectroscopy \citep{vanzella2020}.  \textcolor{black}{\cite{Naidu_2022} indirectly classified $z\sim2$ LAEs into strong LyC leaking and non-leaking galaxies based on the shape and peak separation of their \lya\ profile. The composite spectrum of the potential LyC emitters shows narrow \textrm{C}\textsc{iv} as well as \textrm{He}\textsc{ii}, while there is no sign of these lines in the analogous stack of the non-leakers \citep[see also][]{Saldana-Lopez2022b}.}
If confirmed this link would be really important to probe cosmic reionization: at $z > 5$ directly detecting the Lyman continuum photons escaping from galaxies is impossible because of the IGM opacity \citep{inoue2014}. We therefore need indirect indicators i.e. observational properties that correlate with $f_{esc}$ and can be observed in the epoch of reionization. Many such indirect diagnostics have been proposed \citep[e.g.,][]{izotov2018,Marchi_2018,Verhamme2017}, but all still present a large scatter \citep[see][for a recent comprehensive study on low redshift LyC diagnostics]{flury2022}.

To better interpret this intriguing link between nebular emission and the production and escape of ionizing photons, and the important consequences it could have for reionization, detailed investigations need to be conducted at low and intermediate redshift. However nebular \textrm{C}\textsc{iv} emission is extremely rare in normal star forming (SF) galaxies at least at low and intermediate redshift. \cite{Berg_2019} detected \textrm{C}\textsc{iv} and \textrm{He}\textsc{ii} in two nearby SF galaxies and proposed that this combination of emission may identify galaxies that produce and transmit a substantial number of high-energy photons contributing to cosmic reionization. \cite{Senchyna_2022} also found that both nebular \textrm{C}\textsc{iv} and \textrm{He}\textsc{ii} appear in nearby lower metallicity galaxies ($< 0.1 \ Z_\odot$), provided the specific star formation rate (sSFR) is large enough to guarantee a significant population of high-mass stars, although the equivalent widths (EWs) appear smaller than those measured in individual systems at $z > 6$.
 At intermediate redshift ($z \sim 2-3$), although a hint of \textrm{C}\textsc{iv} emission appears in deep stacked spectra of SF galaxies, especially when combining sub-sample of galaxies with Ly$\alpha$ in emission \citep{Shapley_2003,Pahl_2021}, individual detections are very rare. 
 \cite{Tang_2021} investigated a large sample of more than 100 extreme emission line galaxies, selected to have strong \textrm{O}\textsc{iii}, and detected \textrm{C}\textsc{iv} emission in only 5 sources, but with EWs up to 25 \AA. 
 Similarly at $z\sim2$, \cite{du2020}, analysed a large sample of possible analogs of EoR sources but only found two objects with strong rest-UV metal line emission including significant \textrm{C}\textsc{iv} which appear comparable to the $z > 6.5$ galaxies. Only in low-mass galaxies, which are mostly observable thanks to lensing magnification \citep{Vanzella_2016, Mainali_2017, vanzella2020, Vanzella_2021}, does there seem to be a somewhat higher fraction of sources showing prominent UV emission lines, with the most extreme \textrm{C}\textsc{iii}] emitters, also showing nebular \textrm{C}\textsc{iv} emission with rest EWs 3-8 \AA \ \citep{Stark_2015, Llerena2022}. These galaxies show very large sSFRs indicating that they are in a period of rapid stellar mass growth and have very blue continuum UV slopes indicating minimal dust content, young ages, and low metallicity \citep{Amorin2017}.
 
In this context, the VANDELS survey \citep{McLure_2018,Pentericci_2018} provides the ideal database to search for galaxies showing the rare and faint \textrm{C}\textsc{iv} nebular emission line, since it provides some of the deepest spectra of intermediate redshift star forming galaxies available to date. In particular VANDELS spectra can identify this emission line in galaxies in a redshift range ($\sim 3-4$) where also the direct detection of Lyman continuum flux is possible and thus a confirmation of the link between Lyc escape and \textrm{C}\textsc{iv} emission can be directly tested. With this aim we have thus searched for a sub-sample of galaxies at $z\sim 2.5-5$ that show strong nebular \textrm{C}\textsc{iv} emission from the VANDELS survey, in order to determine their physical proprieties and identify their nature as possible analogs of the cosmic reionizers.
 
The paper is structured as follows. The selection criteria and the sample of \textrm{C}\textsc{iv} emitters in the VANDELS survey are discussed in Sect. \ref{sec2}. Sect. \ref{sec3} examines their spectroscopic properties, while Sect. \ref{sec4} analyzes the emission lines' properties to distinguish star forming galaxies from AGNs in our sample. In Sect. \ref{sec5} we investigate the physical properties of the star forming \textrm{C}\textsc{iv} emitters and in Sect. \ref{sec:LyC_leakers} we investigate how these properties are linked with cosmic reionization physical components. Finally, we summarize our findings in Sect. \ref{sec:conclusions}.

Throughout this paper, we assume a flat $\Lambda$CDM cosmology with $H_0$ = 67.7 km s$^{-1}$ Mpc$^{-1}$ and $\Omega_m$ = 0.307 \citep{refId0}. All magnitudes are expressed in the AB system \citep{Oke_1983}.
\section{The VANDELS \textrm{C}\textsc{iv} emitters}\label{sec2}
\subsection{Sample selection}

For our analysis we have used data from "VANDELS: a VIMOS survey of the CDFS and UDS fields", which is a recently completed ESO public spectroscopic survey carried out using the VIMOS spectrograph on the Very Large Telescope (VLT). VANDELS targets are selected in the \textit{UKIDSS Ultra Deep Survey} (UDS) (RA = 2:18, Dec = -5:10), and the \textit{Chandra Deep Field South} (CDFS) (RA = 3:32, Dec = -27:48): VANDELS footprints are centered on the HST areas observed by the CANDELS program \citep{Grogin_2011,Koekemoer_2011}, although given the VIMOS field of view, they are larger.
The survey description and initial target selection strategies can be found in \cite{McLure_2018}, while data reduction and redshift determination methods can be found in \cite{Pentericci_2018}. Here we give a brief description of the most important aspects relevant to this work.

VANDELS targeted star forming galaxies at $z > 2.4$ as well as massive, passive galaxies in the redshift range $1 < z < 2.5$, with integration times ranging from 20 to 80 hours depending on the magnitude to ensure a uniform signal-to-noise ratio (S/N) on the continuum, with most targets observed for 40 hours or more. The \textsc{EasyLife} data reduction pipeline was used to reduce the spectra \citep{Garilli_2012}, producing the extracted 1D spectra, re-sampled 2D spectra, and sky-subtracted spectra. 
Spectroscopic redshifts were derived using the \texttt{pandora.ez} tool \citep{Pandora_2010} by cross-correlation with galaxy templates. The reliability of the redshifts is quantified with a quality flag (QF) as follows: 0 means no redshift could be determined; 1, 2, 3 and 4 mean respectively a probability of 50\%, 75\%, 95\% and 100\% of the redshift being correct; finally 9 means that the spectrum shows a single emission line. 
The redshift measurement accuracy of spectroscopic observations is typically 150 km s$^{-1}$. For more information, see \cite{Pentericci_2018} for details on data reduction and redshift determination. 

In this work we have used galaxies from the fourth and final data release (DR4) which is described in \cite{Garilli_2021}. The data release contains spectra of approximately 2100 galaxies in the redshift range of $1.0 < z < 7.0$, with over 70 per cent of the targets having at least 40 hours of on-source integration time. 
\subsection{Identification of \textrm{C}\textsc{iv} emitters}
We base our work on the official VANDELS emission line catalog which will be presented in Talia et al. (submitted). Specifically, we use the catalog that was build using Gaussian fit measurements performed with \texttt{slinefit}\footnote{\url{https://github.com/cschreib/slinefit}}\citep{Schreiber_2018}, an automated code that models the observed spectrum of a galaxy as a combination of a stellar continuum model and a set of emission and absorption lines. A set of templates is linearly combined to best fit the continuum. The templates were built with the \cite{Bruzual_2003} stellar population models and fit to the EAzY \citep{Brammer_2008} default template set using FAST \citep{Kriek_2009}. The code searches for lines around their expected locations given by the VANDELS official redshift. Each line with a S/N higher than 3 is allowed its own offset, with a maximum value set by $\pm 1000$ km/s while lines with a lower S/N are instead fixed at their expected position. Therefore, for each detected line, the catalog is formed with the following information: flux and error, rest-frame equivalent width ($EW_0$) and error, and the rest-frame full width at half maximum (FWHM) with error. By using the Monte Carlo technique, these errors are determined by randomly perturbing the galaxy spectrum based on its error spectrum, and then computing the uncertainties from the standard deviation of these perturbations compared to the original data.

For our analysis we selected galaxies with a reliability flag 3, 4 and 9, which show a \textrm{C}\textsc{iv} emission line with a S/N greater than 2.5.

To ensure the reliability of the lowest S/N detections and following \citep{Saxena_2020} we produced a stacked spectrum of the 12 sources with $2.5 \leq S/N \leq 3$, as an additional check to ensure that they are indeed \textit{bona fide} \textrm{C}\textsc{iv} emitters (see also Sect. \ref{sec:stack}). In this way we verified that we can safely consider these sources in the sample.
We selected a total of 50 galaxies.

From this list generated by the automatic procedure we removed by eye six sources which present clear problems of sky-residuals in the region of the \textrm{C}\textsc{iv} emission and one source with evidence of merger in the spectrum. 
 Finally from a visual inspection of all VANDELS QF 3, 4 and 9 sources we note that the official catalog failed to detect some galaxies with clear \textrm{C}\textsc{iv} emission lines which were not correctly identified by \texttt{slinefit}. This is probably due to the limited velocity offset allowed in the automatic search and to the fact that the reference redshift used for the search is the "VANDELS redshift", which is not the systemic redshift but is in general driven by the strongest features present in the spectrum. In few cases this redshift is closer to the Ly$\alpha$ redshift or to the ISM redshift depending on their relative strength. A further 8 sources were identified and re-analyzed through \texttt{slinefit} by correctly calibrating the redshift (on the \textrm{C}\textsc{iii}] line when possible) which represents better the systemic redshift of the galaxies. 
 
\subsection{Removal of X-ray AGN/BLAGN and final sample}

Since we want to study the nature of \textrm{C}\textsc{iv} in star forming galaxies, we excluded the sources which are obvious AGN: in particular we excluded 8 X-ray detected sources from the CDFS 7 Ms Source Catalogs \citep{Luo_2017} and the UDS X-ray observations \citep{Kocevski2018}, of which 4 are broad line AGNs (\textit{BLAGNs}: CDFS003360, CDFS019824, UDS017629 and UDS199159) which are easy to recognise from the broad UV emission line, and 4 narrow line AGNs (\textit{NLAGNs}: CDFS005827, CDFS019505, UDS025482 and UDS018960). For an individual analysis of these sources see Bongiorno et al. (in prep.). 

In conclusion, from the parent sample made of 933 galaxies at $2.5 \leq z \leq 5$ without the AGN contribution, we obtain a sub-sample of 43 sources, 20 from the UDS field, and 23 from the CDFS field. The IDs, RA and DEC coordinates, redshifts, related flags and emission line information are presented in Table \ref{Tab_prop}. This sample might still contain some NLAGNs with no X-ray counterpart.
In the following section we will refine our analysis and employ emission line ratio diagrams to obtain a sample of purely star forming \textrm{C}\textsc{iv} galaxies.

\begin{figure}
 \centering
 \includegraphics[width=0.45\textwidth]{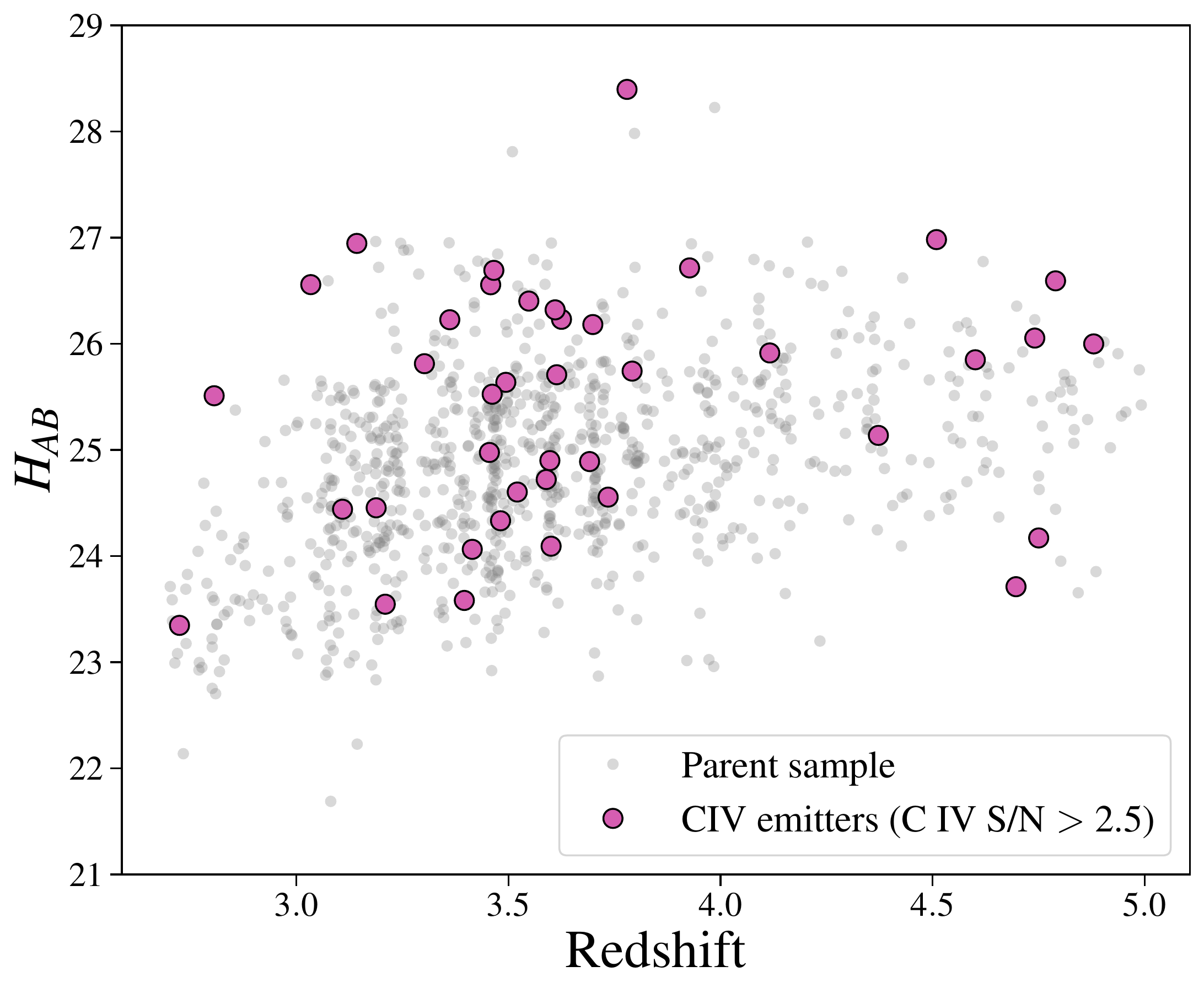}
 \caption{Distribution of $H_{AB}$ of the \textrm{C}\textsc{iv} emitters with redshift. The sources cover a redshift range of $2.8 < z < 5$, with a median redshift of $\bar{z} \sim $ 3.54. Of the 43 selected sources, 3 of them do not show any detection in the H (HST/F160W) band, so they are not shown in the figure.}
 \label{fig:H_z_dist}
\end{figure}

In Fig. \ref{fig:H_z_dist} we show the distribution of the galaxies' $H_{AB}$ magnitude as a function of the redshift for the \textrm{C}\textsc{iv} sample (highlighted in purple) and the parent sample (grey symbols), i.e. all other VANDELS sources after removing the AGNs, as classified in Bongiorno et al. (in prep.). For this comparison the parent sample has been cut between redshift 2.5 and 5. As we can see the \textrm{C}\textsc{iv} emitters are a representative subset of the VANDELS population at least in terms of H-band and redshift distribution.

\begin{table*}
 \caption{Main features of the \textrm{C}\textsc{iv} emitters. \textit{AGN} indicates the nature of the sources: $\checkmark$ for X-ray/BLAGN, ? for a possible AGN and $\times$ for a purely star forming galaxy. The measured \textrm{He}\textsc{ii}, \textrm{C}\textsc{iii}], \textrm{O}\textsc{iii}] and \textrm{C}\textsc{iv} rest-frame line fluxes are in units of $\times 10^{-18} erg \ s^{-1} \ cm^{-2}$, and the $EW_0$ of the \textrm{C}\textsc{iv} line is un units of \AA. The sources with the asterisk are the ones re-analyzed with \texttt{slinefit} and not contained in Talia et al. (submitted). The sources in red are the best candidates to be strong LyC leakers (as discussed in Sec. \ref{sec:LyC_leakers}).}
  \label{Tab_prop}
 
 $$ 
  \begin{array}{ccclcccccc}
  \hline \hline
  \noalign{\smallskip}
  \textsc{ID} & \textsc{DEC [J2000]}& \textsc{RA [J2000]}& z & AGN & \textrm{He}\textsc{ii} & \textrm{C}\textsc{iii}] & \textrm{O}\textsc{iii}] & \textrm{C}\textsc{iv} & \textsc{$EW_0$ [\AA]} \\
  \noalign{\smallskip}
  \hline
  \noalign{\smallskip}
  \textcolor{red}{CDFS000303} & -27.937461	& 53.069643 & 3.6137 & \times & < 0.6 & 1.2 \pm 0.3 & < 1.2 & 1.0 \pm 0.2 & 3.7 \pm 0.9 \\
  CDFS003839 & -27.875749	& 53.144602 & 3.458 & \checkmark & 1.6 \pm 0.6 & < 1.2 & < 2.2 & 3.4 \pm 0.8 & 14 \pm 3\\
  CDFS006327 & -27.850722	& 53.178278 & 3.4937 & \checkmark & 1.4 \pm 0.4 & 1.0 \pm 0.4 & < 0.6 & 1.7 \pm 0.4 & 7.8 \pm 1.7\\
  \textcolor{red}{CDFS012637} & -27.795843	& 53.122588& 3.4655 & ? & < 0.8 & 1.3 \pm 0.4 & < 1.8 & 1.6 \pm 0.5 & 6 \pm 2 \\
  CDFS014277 & -27.785081	& 53.231559 & 3.3616 & \times& < 1.2 & 4.4 \pm 1.5 & < 1.8 & 5.9 \pm 0.7 & 9.8 \pm 1.3\\
  \textcolor{red}{CDFS021733} & -27.724985	& 53.055324 & 3.1087 & \times & < 1.7 & < 3 & < 1.9 & 6.2 \pm 1.4 & 29 \pm 8 \\
  CDFS021776 & -27.724619	& 53.098507 & 4.741 & \times& < 2.0 & - & < 4.0 & 2.6 \pm 0.8 & 13 \pm 5\\
  \textcolor{red}{CDFS025178} & -27.689563	& 53.184370 & 2.806 & ? & < 1.4 & < 1.6 & < 1.9 & 4.4 \pm 0.9 & 5.1 \pm 1.0\\
  CDFS025760 & -27.696507	& 53.069094 & 4.79 & \times& 92 \pm 5 & - & 67 \pm 14 & 32 \pm 3 & 20 \pm 2 \\
  CDFS028933 & -27.848752	& 53.029622 & 3.7795 & \times& < 1.0 & 1.9 \pm 0.3 & < 1.0 & 2.2 \pm 0.2 & 9.4 \pm 1.0\\
  CDFS029763 & -27.825989	& 53.238443 & 3.105 & ?& < 0.4 & < 0.5 & 0.8 \pm 0.2 & 0.7 \pm 0.3 & 8 \pm 3\\
  CDFS032490 & -27.754031	& 53.219545 & 3.0747 & ?& < 0.6 & 1.1 \pm 0.4 & < 0.4 & 1.1 \pm 0.4 & 9 \pm 3\\
  \textcolor{red}{CDFS128455} & -27.745586	& 53.003635 & 3.4813 & \times& < 1.5 & 3.9 \pm 1.2 & < 2.1 & 7 \pm 3 & 3.5 \pm 1.4 \\
  CDFS202937 & -27.963161	& 53.218947 & 3.735 & \checkmark & 4.2 \pm 0.6 & 3.1 \pm 0.8 & < 2.1 & 4.3 \pm 0.6 & 8.9 \pm 1.3\\
  CDFS203861 & -27.956412	& 53.215551 & 4.7502 & \times & < 5.0 & - & < 3.8 & 5.0 \pm 1.7 & 19 \pm 7 \\
  \textcolor{red}{CDFS203989} & -27.955446	& 53.019099 & 3.1366 & \times& 1.4 \pm 0.4 & 4.1 \pm 0.5 & 1.8 \pm 0.3 & 4.5 \pm 0.7 & 8.8 \pm 1.4\\
  CDFS208175 & -27.923411	& 53.266337 & 3.462 & ?& < 1.4 & 1.1 \pm 0.4 & < 1.8 & 3.6 \pm 1.2 & 3.8 \pm 1.3 \\
  CDFS209144 & -27.916473	& 53.070682 & 4.4286 & \times& < 2.0 & - & < 1.9 & 2.5 \pm 0.9 & 8 \pm 3.1\\
  \textcolor{red}{CDFS212043} & -27.894033	& 52.994478 & 3.395 & \times& < 1.6 & 4.2 \pm 0.8 & < 1.4 & 2.9 \pm 0.9 & 3.4 \pm 1.1\\
  \textcolor{red}{CDFS231741} & -27.747690	& 53.287112 & 4.0068 & \times & < 0.9 & 3.6 \pm 1.2 & 2.8 \pm 0.6 & 2.9 \pm 0.7 & 6.3 \pm 1.4 \\
  CDFS247279 & -27.676973	& 53.206236 & 3.2959 & \times& < 1.8 & 4.0 \pm 0.6 & < 2.6 & 4.3 \pm 0.8 & 7.4 \pm 1.5\\
  CDFS009384^\star & -27.823526	& 53.078521 & 3.5480 & \times& 0.8 \pm 0.2 & <1 .6 & < 1.5 & 0.33 \pm 0.11 & 1.1 \pm 0.4 \\
  \textcolor{red}{CDFS025002^\star} & -27.688314	& 53.066713 & 3.7915 & ?& < 1.6 & 1.7 \pm 0.8 & < 19 &1.9 \pm 0.5 & 4.4 \pm 1.3\\
  \textcolor{red}{UDS000184} & -5.275796 & 34.489558 & 4.3722 & \times& < 2.7 & < 15 & 6.6 \pm 1.1 & 6.5 \pm 1.3 & 10 \pm 2\\
  \textcolor{red}{UDS003724} & -5.255394 & 34.395118 & 3.4616 & \times& < 1.3 & 3.8 \pm 0.9 & < 4.4 & 4.5 \pm 0.7 & 7.6 \pm 1.2\\
  UDS018872 & -5.171582 & 34.371472 & 4.8799 & \times& < 3.7 & - & < 5.4 & 3.3 \pm 1.1 & 10 \pm 4\\
  UDS020394 & -5.163315 & 34.535879 & 3.3018 & \times& < 1.1 & 3.9 \pm 0.4 & < 5.0 & 2.5 \pm 0.8 & 5.1 \pm 1.6\\
  UDS021234 & -5.158691 & 34.420972 & 4.6008 & \times& 2.2 \pm 0.6 & - & < 15 & 5.7 \pm 1.0 & 17 \pm 4 \\
  UDS021291 & -5.158290 & 34.295009 & 3.625 & \times& < 0.9 & 2.6 \pm 1.0 & < 1.3 & 1.9 \pm 0.6 & 5.5 \pm 1.9\\
  UDS033733 & -5.162411 & 34.406837 & 3.1424 & \times& < 0.8 & < 1.5 & < 0.6 & 1.4 \pm 0.5 & 12 \pm 5\\
  UDS034251 & -5.153870 & 34.581079 & 3.61 & ?& < 1.3 & < 1.6 & < 1.4 & 2.1 \pm 0.8 & 6 \pm 2\\
  UDS034333 & -5.152631 & 34.404969 & 3.927 & \times & < 0.8 & < 2.1 & < 2.0 & 2.5 \pm 0.6 & 14 \pm 4\\
  UDS145830 & -5.336489 & 34.556466 & 3.2094 & \checkmark & 8.3 \pm 1.0 & 3.5 \pm 1.2 & < 3.4 & 5.5 \pm 1.3 & 18 \pm 6\\
  UDS283510 & -5.368315 & 34.388651 & 4.5092 & \times& < 2.7 & - & < 3.7 & 5.3 \pm 1.1 & 18 \pm 5\\
  UDS313619 & -5.287081 & 34.242560 & 3.188 & \times& < 2.4 & 10.7 \pm 1.8 & 4.6 \pm 1.7 & 6.6 \pm 1.9 & 7 \pm 2\\
  UDS379641 & -5.110200 & 34.334556 & 4.1161 & \times& < 1.8 & < 4.7 & < 3.5 & 2.9 \pm 1.0 & 9 \pm 3 \\
  UDS395442 & -5.070635 & 34.434154& 3.589 & \times& < 1.3 & < 2.5 & < 1.1 & 2.6 \pm 0.9 & 2.3 \pm 0.8\\
  UDS020932^\star & -5.160535 & 34.518914 & 3.6005& \times& < 2.4& < 10 & < 3.1 & 3.2 \pm 0.9& 3.5 \pm 1.0\\
  UDS140603^\star & -5.361135 & 34.393002 & 3.4146 & ?& < 3.0 & < 4.2 & < 3.0 & 5.0 \pm 1.0 & 5.6 \pm 1.2 \\
  \textcolor{red}{UDS152744^\star} & -5.303281 & 34.393005 & 2.7243 & ?& < 3.2& < 1.9 & < 2.4 &5.2 \pm 1.1 & 2.4 \pm 0.5 \\
  UDS312746^\star & -5.289354 & 34.243205 & 3.5207 & \times& < 2.1& < 6.8 & < 5.4 & 2.2 \pm 0.7 & 3.2 \pm 1.1\\
  UDS315235^\star & -5.282526 & 34.473701 & 3.5976 & \times& < 2.4& < 4.1 & < 4.4 & 1.8 \pm 0.6 & 2.6 \pm 0.9\\
  UDS388069^\star & -5.089615 & 34.339181 & 3.4550 & ?& < 1.7 & < 2.0 & < 1.1 & 2.3 \pm 0.4 & 3.7 \pm 0.7 \\
  \noalign{\smallskip}
  \hline
  \end{array}
 $$ 
 \end{table*}
 
 \section{Spectroscopic properties of the \textrm{C}\textsc{iv} emitters}\label{sec3}

\begin{figure}
 \centering
 \includegraphics[width=0.45\textwidth]{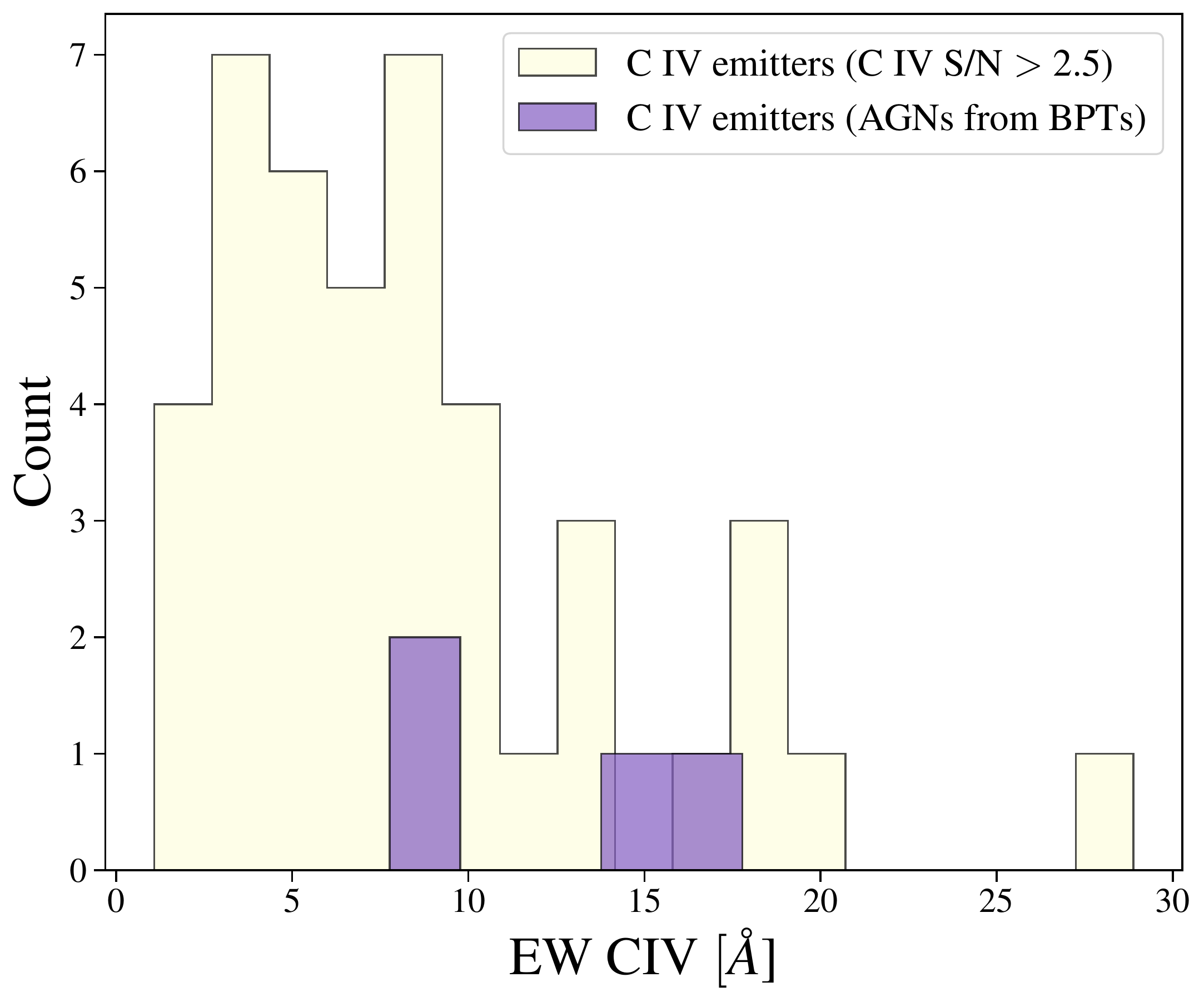}
 \caption{Light distribution: measured $EW_0$ of the \textrm{C}\textsc{iv} emission line for the \textrm{C}\textsc{iv} emitters (S/N > 2.5) identified in both CDFS and UDS fields. Purple distribution: measured $EW_0$ of the \textrm{C}\textsc{iv} emission line for the sample of certain AGNs described in Sec. \ref{sec:AGN_SF}.}
 \label{fig:CIV_dist}
\end{figure}

\begin{figure*}
 \centering
 
 \subfloat{\includegraphics[width=0.5\textwidth]{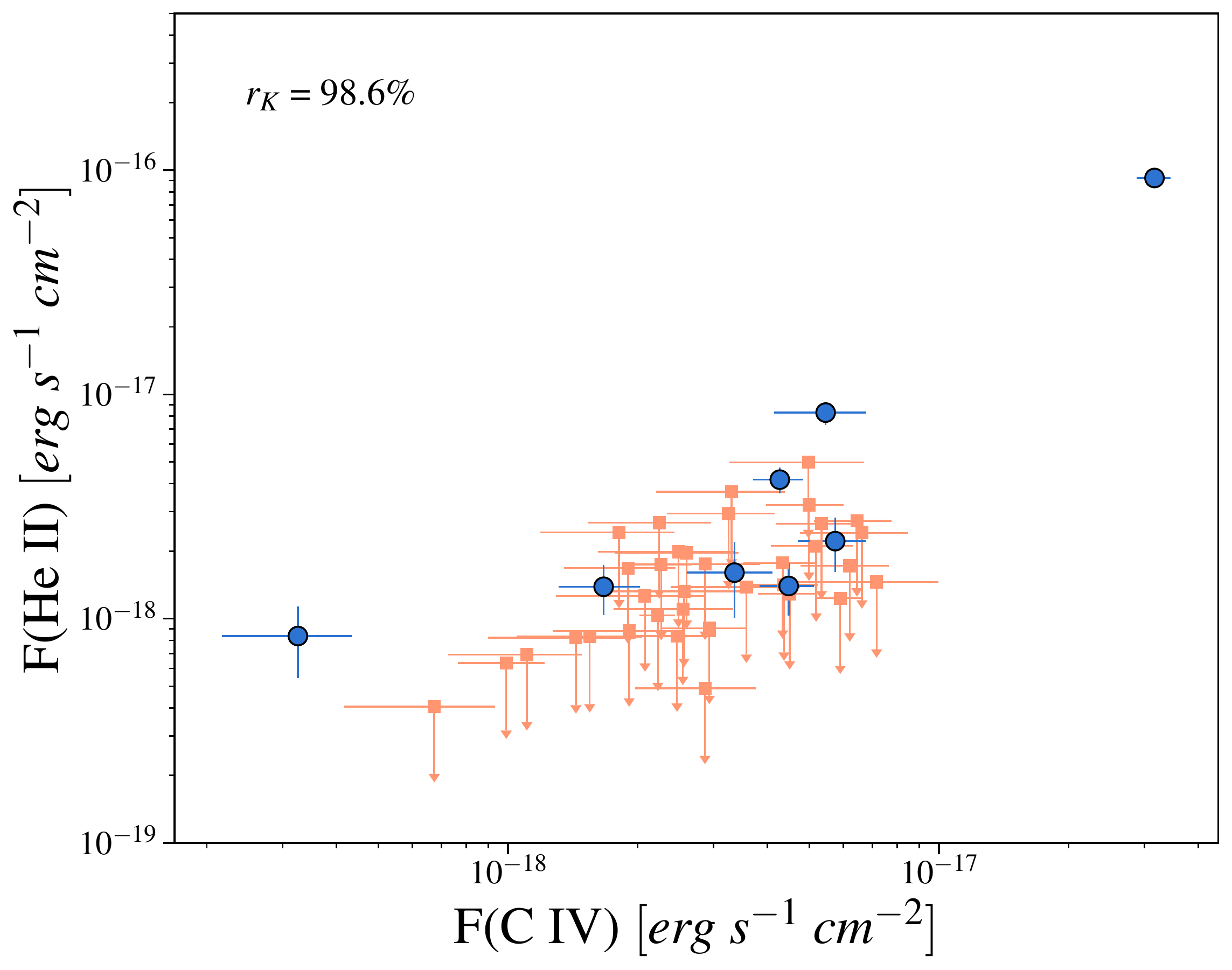}}
 \subfloat{\includegraphics[width=0.5\textwidth]{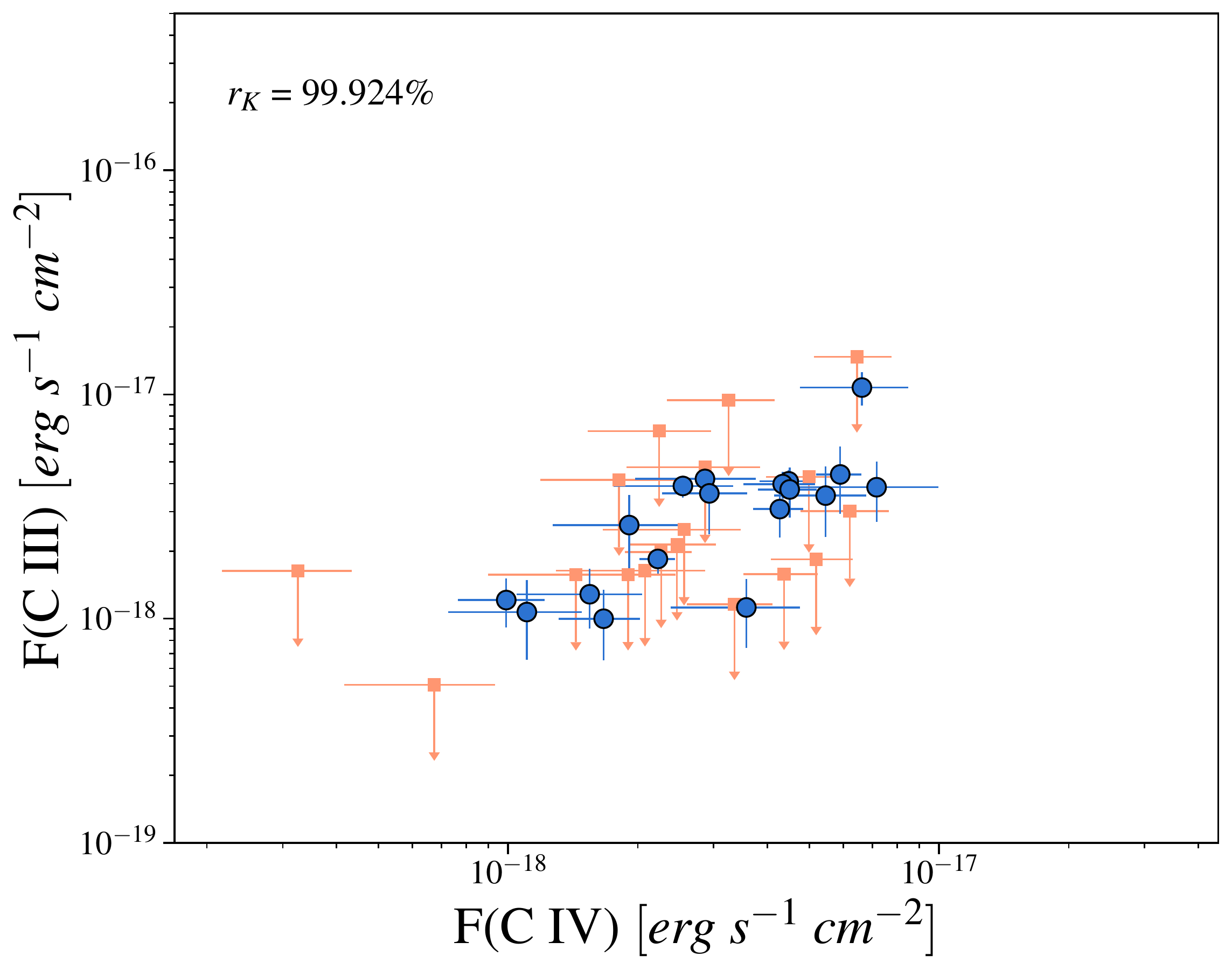}}\quad
 \subfloat{\includegraphics[width=0.5\textwidth]{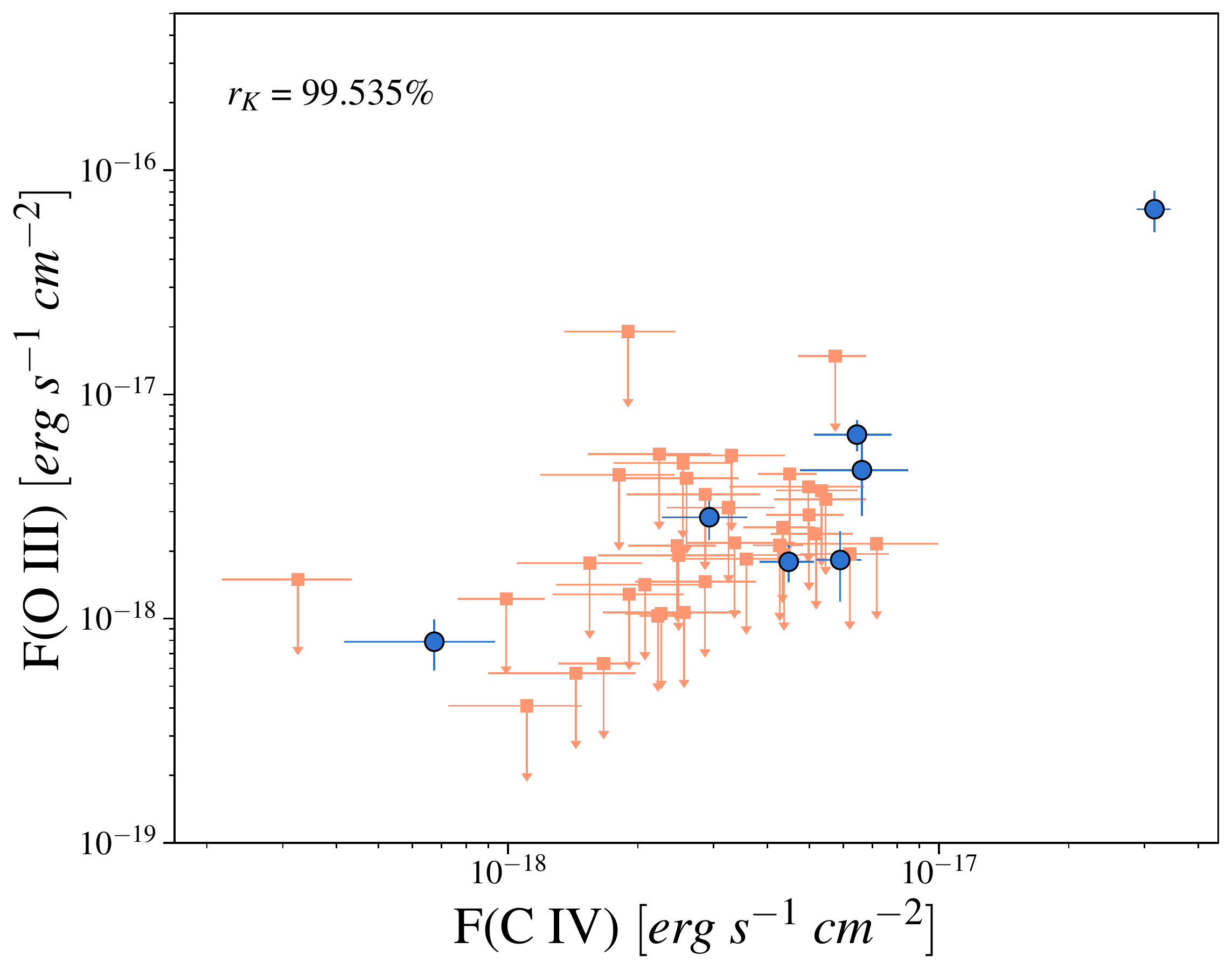}}
 \subfloat{\includegraphics[width=0.5\textwidth]{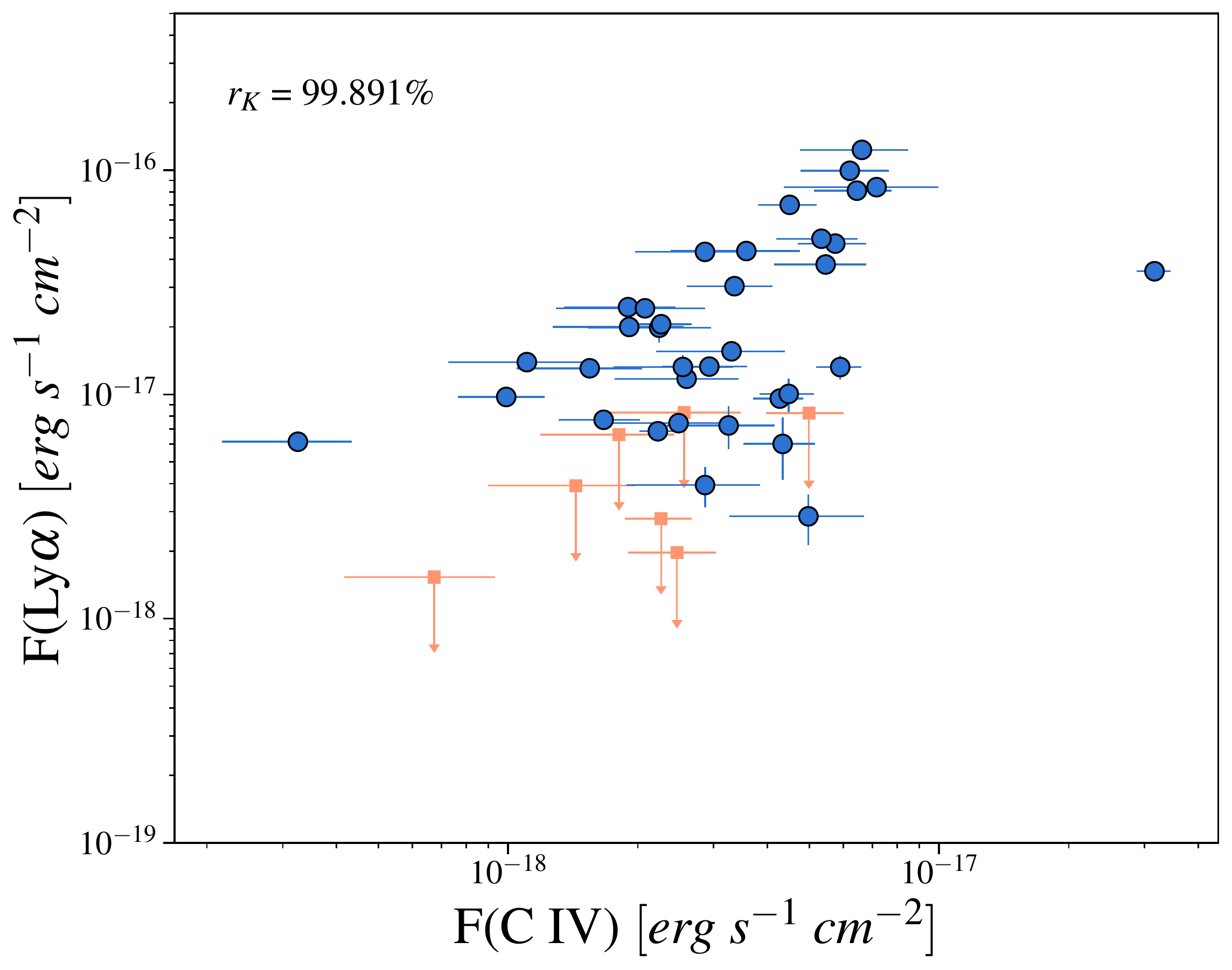}}\quad
 \caption{Distribution of the measured \textrm{He}\textsc{ii}, \textrm{C}\textsc{iii}], \textrm{O}\textsc{iii}] and Ly$\alpha$ line fluxes and the \textrm{C}\textsc{iv} line flux for the selected \textrm{C}\textsc{iv} emitters ($S/N > 2.5$) identified in both CDFS and UDS fields (blue points). If $S/N$ of a line is lower than 2.5, we set 2$|\sigma|$ as upper limit (orange squares). In the figure it is also reported the correlation coefficient $r_K$ from the generalized Kendall's $\tau$ statistic. 
 }
 \label{fig: line_comparison}
\end{figure*}

 In Fig. \ref{fig:CIV_dist} we present the distribution of the rest-frame equivalent widths ($EW_0$s) of the \textrm{C}\textsc{iv} emission line in our sample, which range from 1 to 29 \AA. Therefore, although most objects show $EW_0$ of a few \AA, in the VANDELS dataset we also find some galaxies with extreme EWs, analogous to those found in the epoch of reionization \citep{stark2014,Mainali_2018}.
In Table \ref{Tab_prop} we report the 
 fluxes of the \textrm{C}\textsc{iv} lines and their $EW_0s$. 
We also present the fluxes for the \textrm{He}\textsc{ii}, \textrm{O}\textsc{iii}] and \textrm{C}\textsc{iii}] emission lines from the catalog of Talia et al. (submitted) plus the 8 additional sources considered in this work. We then analyse the relation between the \textrm{C}\textsc{iv} emission line and the other UV lines and we present the results in Fig. \ref{fig: line_comparison}, where we plot as limits all lines detected with a S/N lower than 2.5.
We see a good correlation in all cases. 
In the insets of Fig. \ref{fig: line_comparison} we also indicate, for each distribution, the probability $r_K$ that the two variables are dependent, following the generalized Kendall's $\tau$ statistic including the upper limits \citep{Isobe1986}.
Other studies at similarly high redshift have investigated analogous relations between UV lines, although only focusing on the \textrm{C}\textsc{iii}] emission line: for example \cite{Schmidt_2021} report that the strengths of \textrm{He}\textsc{ii}, \textrm{O}\textsc{iii}, and \textrm{Si}\textsc{iii} correlate with the flux of the \textrm{C}\textsc{iii}] emission while \cite{Marchi_2019, Cullen_2020} found that that the Ly$\alpha$ EW and the \textrm{C}\textsc{iii}] EW are strongly related \citep[see also][]{Stark_2014} although this relation has a really large scatter \citep[e.g.,][]{Lefevre_2019} and therefore it appears strong only when using stacks.

 \textcolor{black}{Even if our emission lines are all quite close in wavelength, to perform a quantitative analysis, dust reddening should be consider. Due to the lack of direct measurements of \ha\ and \hb, we used the SED fitting results (see Sec. \ref{sec:beagle}) to estimate the amount of corrections for dust attenuation. From the stellar $A_v\text{(SED)}$ we computed the stellar reddening $E(B-V)$ and then the nebular $E(B-V)$ \citep[following][]{Calzetti2000}: this quantity is on average 0.11. These results are consistent with those reported by \cite{Saxena_2022}, in which a $E(B-V)_{nebular} \sim 0$ was obtained analyzing the stack of the brightest sub-sample of the VANDELS \civalone\ emitters presented in this work. Since the dust correction inferred would only marginally affect our line ratios, we present only observed values.}

 \textcolor{black}{For the \civalone\ emission we must also consider a correction for a component due to stellar emission.
Stellar \civalone\ produces a classical p-Cygni profile, that is present only in a small subset of our sources.
 We consider the results of \cite{Saxena_2022}, in which we derived a stellar component to be less than 25\% on the strong \civalone\ by fitting the individual galaxy spectra using SEDs that only contain stellar emission as done in \cite{Saldana-Lopez_2022}. As we will show in the next Section, the p-Cygni profile is missing in the stacked spectrum of all SF galaxies, again indicating that the stellar component is probably very small on average in the total sample and possibly present only in the stronger emitters. 
In the following analysis we will use the uncorrected \civalone, but we will indicate in the various cases if and how this correction could alter our results.}

\section{Classification of SF/AGN based on UV emission lines}\label{sec4}
None of the 43 sources in our sample have X-ray individual detections: however some of them might still be faint AGN. In order to identify only the population of star forming galaxies, we compare the UV emission properties of our sources to the predictions of photoionization models of AGN and star forming galaxies. At variance with the seminal work by \citet{BPT_1981} who first produced diagrams to discriminate AGN from star forming galaxies, we use UV emission lines instead of those falling in the optical rest frame.

Specifically we consider the \textrm{C}\textsc{iv}, \textrm{He}\textsc{ii} and \textrm{C}\textsc{iii}] (we do not include the \textrm{O}\textsc{iii} emission since it is only detected in few galaxies).
Indeed, these lines are among the most commonly detected UV emission lines in galaxy spectra and they are used to discriminate between photoionization by AGN and shocks in galaxies as well as to investigate the ISM metallicity. The \textrm{C}\textsc{iv}/\textrm{He}\textsc{ii} or \textrm{C}\textsc{iii}]/\textrm{He}\textsc{ii} ratios are sensitive to metallicity, while \textrm{C}\textsc{iii}]/\textrm{C}\textsc{iv} depends mostly on the ionisation parameter \citep[e.g.,][]{Groves_2004, Nagao_2006, Nanayakkara_2019A}. As we will discuss in the next sections, the models of AGN and star forming galaxies occupy two distinct regions in the UV diagnostic diagrams (henceforth also called UV-BPTs for simplicity), so these can be used as discriminators to effectively classify our sources. 

\subsection{AGN models: \cite{Feltre_2016}}\label{sec:AGNmodel}

 The models developed by \citet{Feltre_2016} for the emission by the AGN narrow-line regions have been created using \textsc{Cloudy} \citep{Cloudy_2013}\footnote{These models can be found at \url{http://www.iap.fr/neogal/agn-models.html}}.
 They consider a parameterization of the gas distribution as a cloud of a single type but contaminated by dust and dominated by the radiation pressure. 
It is useful to highlight that in these models, the line measurements are not corrected for dust attenuation. 

For the purposes of our analysis, we use models with $Z$ = 0.0001, 0.0002, 0.0005, 0.001, 0.002, 0.004, 0.006, 0.02 with no priors on $n_H$ in order to identify a clear AGN region in all the UV-BPTs.  \textcolor{black}{These models were constructed using the relative abundances of heavy elements for solar metallicity \citep{Caffau2011}.} The predicted intensities of 20 optical and ultraviolet emission lines are then produced (in units of erg s$^{-1}$). We consider the \textrm{C}\textsc{iv}, \textrm{He}\textsc{ii} and \textrm{C}\textsc{iii}] predicted intensities.
Since we also want to produce a diagnostic diagram of \textrm{C}\textsc{iv} EW with \textrm{C}\textsc{iv}/\textrm{He}\textsc{ii} which was already used by \citep{Nakajima_2018}, we need to also derive the EW of the \textrm{C}\textsc{iv} line from the luminosities provided by the models. To determine the line continuum we followed the procedure outlined in \cite{Castellano_2022}. We first
re-normalize the incident spectra to a chosen fraction of the observed average flux in the R band, i.e. 5950-7250 \AA, which corresponds to the observable non-ionizing UV flux at $\sim$ 1500 \AA, and then compute the EW of the \textrm{C}\textsc{iv} line on the basis of the predicted line flux for each model and of the observed flux at the relevant
wavelength.

\subsection{SF models: \cite{Xiao_2018} and \cite{Gutkin_2016}}

It has been argued that the inclusion of interacting binary stars in stellar population synthesis models may hold the key
to explain the observations of bright high ionization nebular emission lines in the spectra of high redshift star forming galaxies \citep{Stanway_2016, Eldridge_2017}.
\cite{Xiao_2018} combine the output of the stellar spectral synthesis model \textsc{bpass} (Binary Population and Spectral Synthesis) used in conjunction with the photoionization code \textsc{Cloudy}
to predict the nebular emission from \textsc{H ii} regions around young stellar populations over a range of compositions and ages. 
 
In total, there are 80262 photoionization models, including separate models for binary and single stars\footnote{These models can be found at \url{https://bpass.auckland.ac.nz/4.html}}. For our analysis, we consider the photoionization models for binary stars with $Z$ = 0.001, 0.002. 0.02 and $\log(\text{Age}/\text{yr})$ = 6, 6.5, 7, 7.5. For each model, the predicted luminosities and EWs of 17 optical and ultraviolet emission-lines are given (in units of erg s$^{-1}$ and \AA, respectively). We retrieved the \textrm{C}\textsc{iv}, \textrm{He}\textsc{ii} and \textrm{C}\textsc{iii}] predicted intensities and EWs.

In addition, we also consider the models developed by \citet{Gutkin_2016} for the nebular emission from star forming galaxies. They have been created using \textsc{Cloudy} \citep{Cloudy_2013}.
For our analysis, we use models with $Z$ = 0.0001, 0.0002, 0.001, 0.002 and 0.02 with no priors on $n_H$  \textcolor{black}{and with relative elemental abundances fixed
at Solar ratios \citep{Caffau2011}.} The predicted intensities of 20 optical and ultraviolet emission-lines are then produced (in units of erg s$^{-1}$)\footnote{These models can be found at \url{http://www.iap.fr/neogal/sf-models.html}}. As in the previous cases, we consider the \textrm{C}\textsc{iv}, \textrm{He}\textsc{ii} and \textrm{C}\textsc{iii}] predicted intensities.

Similarly to what done with the \cite{Feltre_2016} models, we derive also the EW of the \textrm{C}\textsc{iv} line from the luminosities provided by the models, employing the method described in Sec. \ref{sec:AGNmodel} and in \cite{Castellano_2022}. 

\subsection{Separating SF galaxies from AGN in the UV-BPTs}\label{sec:AGN_SF}

We now compare the properties of our sample of \textrm{C}\textsc{iv} emitters with the predictions from photo-ionisation models for nebular emission from star forming galaxies or AGN narrow-line regions described in the previous sections. 
Following the original diagnostics proposed by \cite{Feltre_2016} we will study the two relations \textcolor{black}{ \textrm{C}\textsc{iii}]/\textrm{He}\textsc{ii} vs \textrm{C}\textsc{iv}/\textrm{He}\textsc{ii} and 
\textrm{C}\textsc{iv}/\textrm{He}\textsc{ii} vs \textrm{C}\textsc{iv}/\textrm{C}\textsc{iii}]}. 
In addition to these, and following the work by \cite{Nakajima_2018}, who constructed photoionization models to interpret the UV spectra of \textrm{C}\textsc{iii} emitters selected from the VUDS survey at $z=2-4$,
we will also use the relation \textrm{C}\textsc{iv}/\textrm{C}\textsc{iii}] versus (\textrm{C}\textsc{iii}]+\textrm{C}\textsc{iv})/\textrm{He}\textsc{ii}, and \textrm{C}\textsc{iv} EW vs \textrm{C}\textsc{iv}/\textrm{He}\textsc{ii}.

\begin{figure*}
 \centering
 \subfloat{\includegraphics[width=0.5\textwidth]{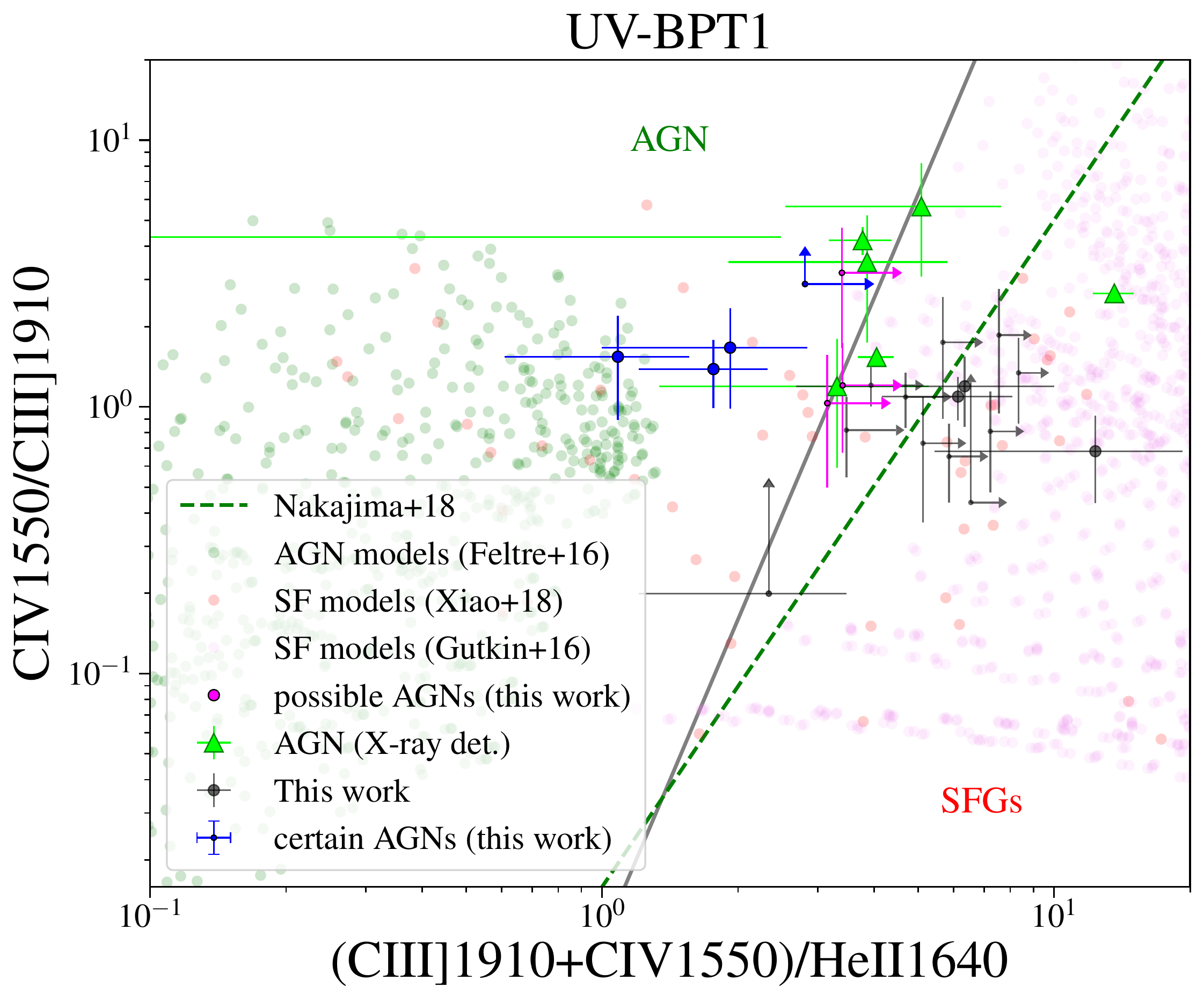}}
 \subfloat{\includegraphics[width=0.5\textwidth]{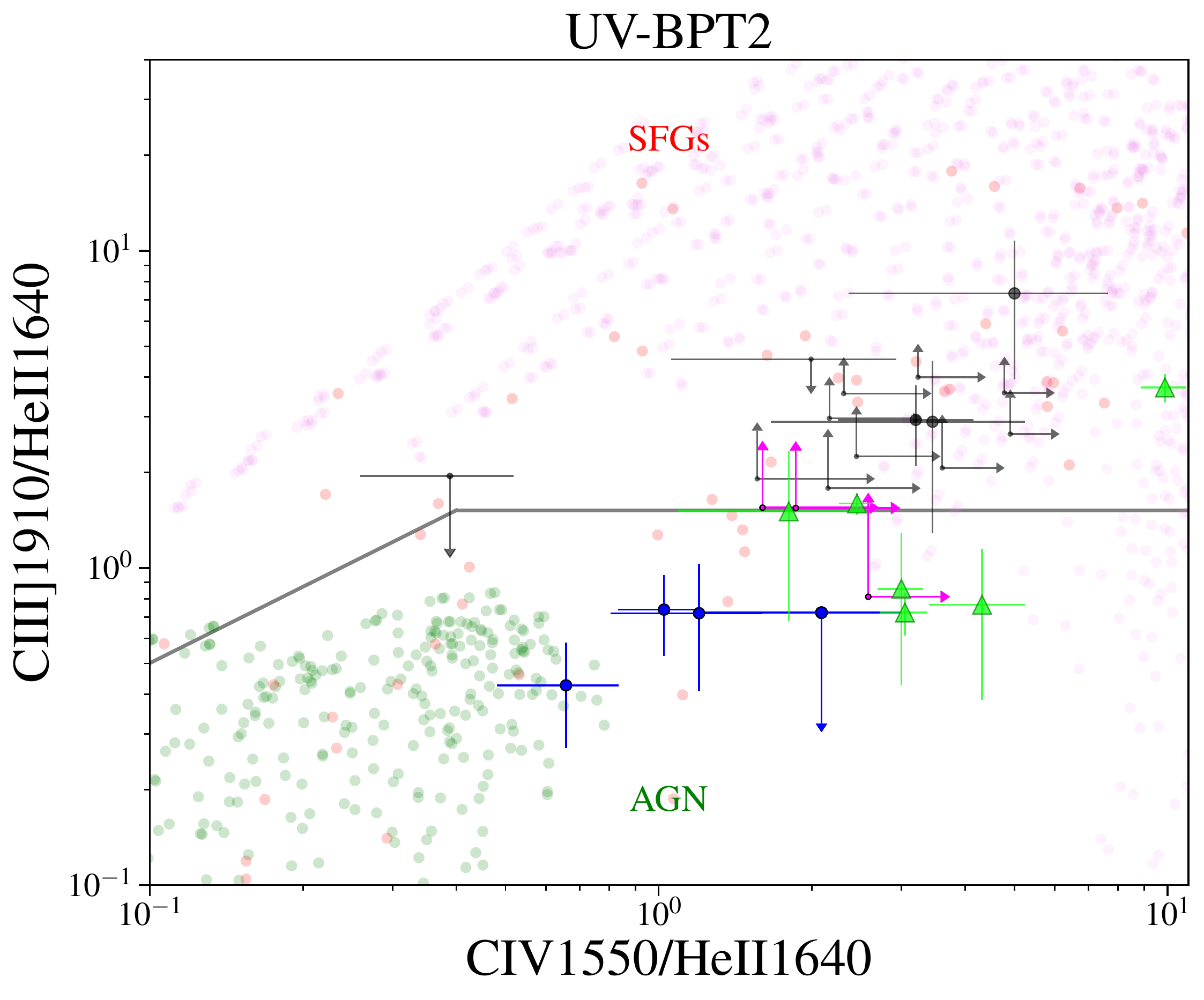}}\quad
\subfloat{\includegraphics[width=0.5\textwidth]{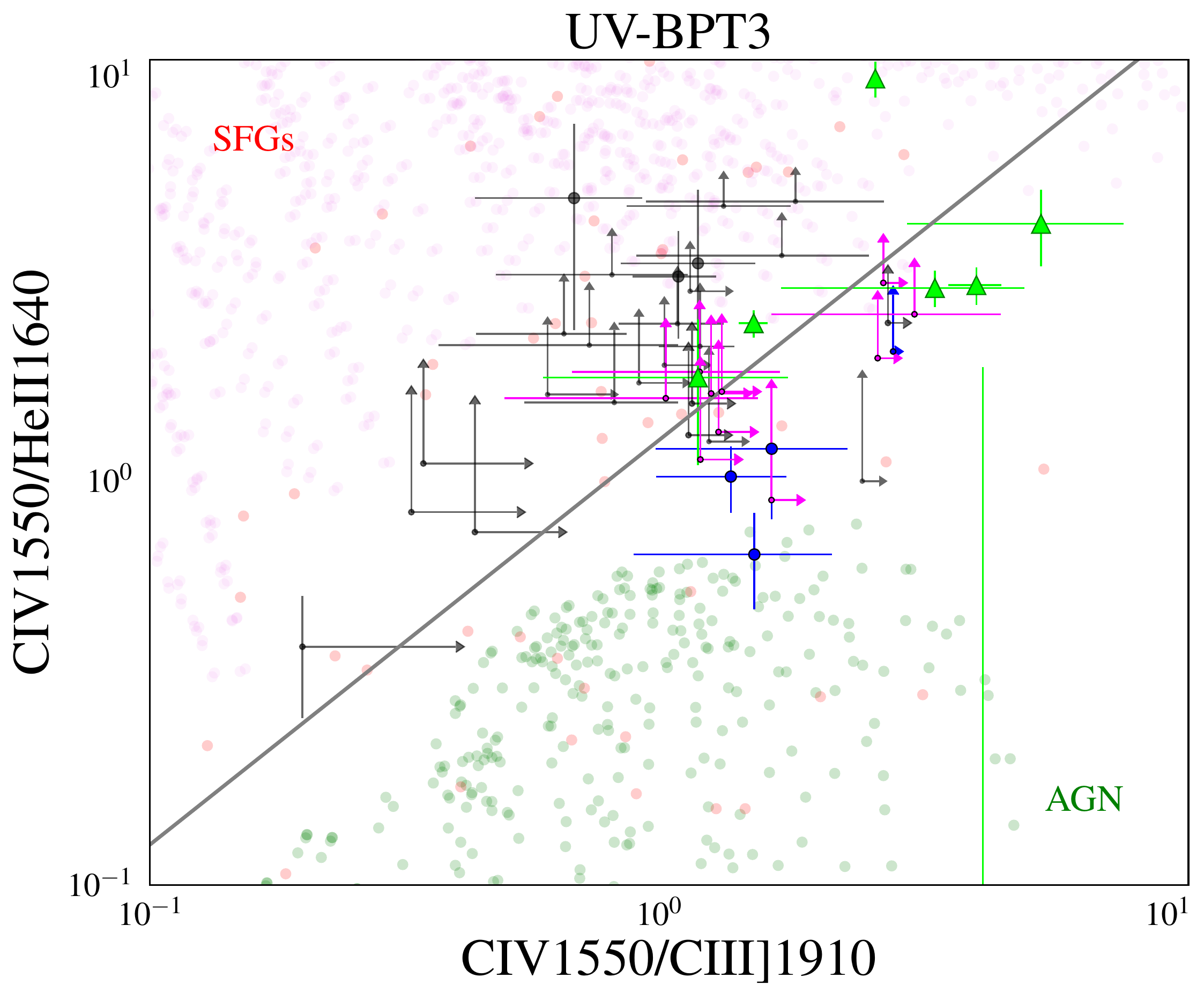}}
\subfloat{\includegraphics[width=0.51\textwidth]{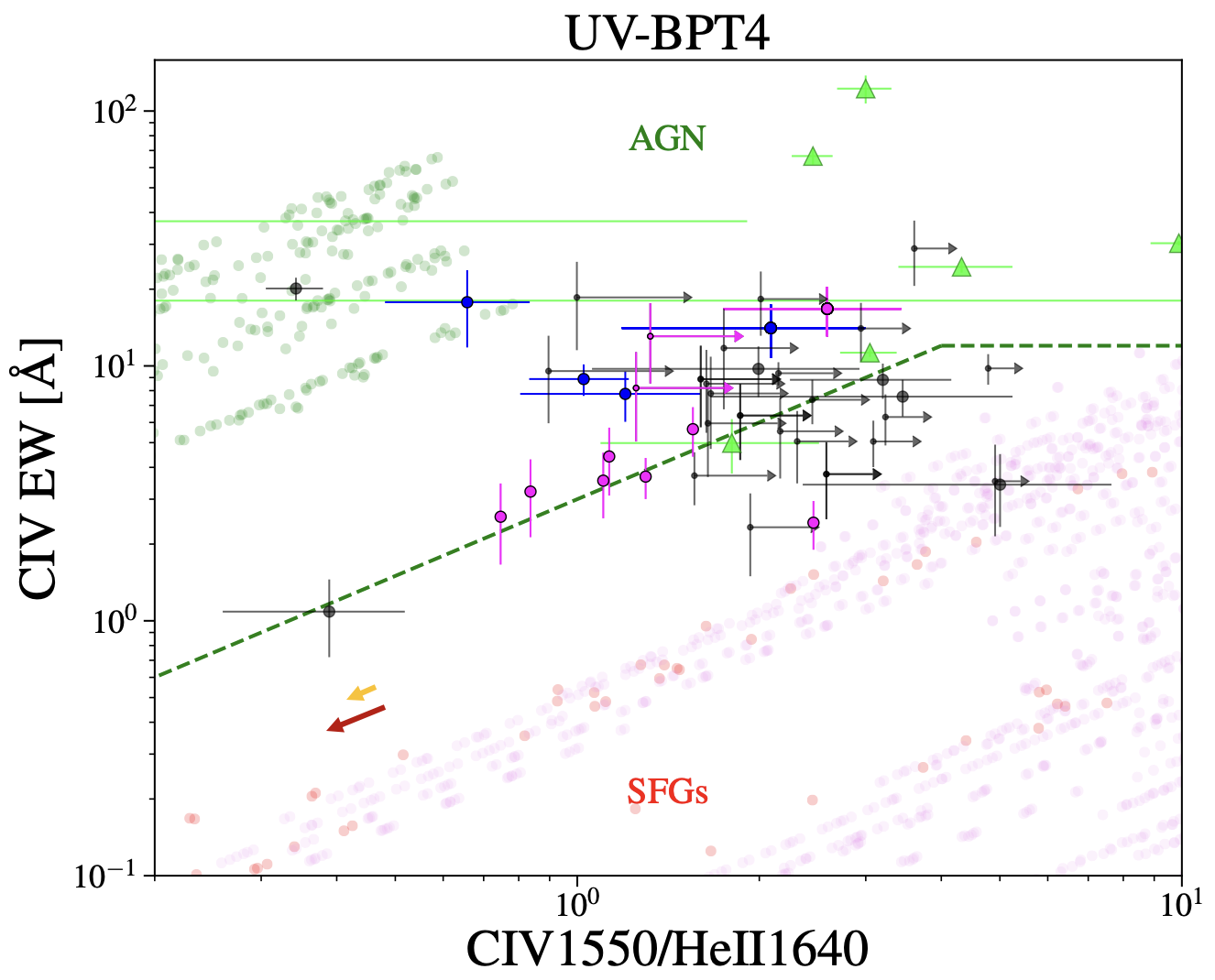}}\quad
 \caption{UV diagnostic diagrams presenting photoionization models of AGN \citep[dark green dots,][]{Feltre_2016} and SF galaxies (SFGs) \citep[red dots,][]{Xiao_2018, Gutkin_2016}. The separating lines consistently divide the two populations, and thus our sample of \textrm{C}\textsc{iv} emitters in: SF galaxies (black dots), confirmed AGNs (X-ray detected, in green), certain AGNs (blue dots) and possible AGNs (magenta), classified according to our criteria. The dashed green line division between ionisation due to star formation alone and ionisation due to AGN has been presented in \cite{Nakajima_2018}. The grey solid line represents our best guess of the possible division between the two regions in all the UV-BPTs.  \textcolor{black}{In the UV-BPT4 diagram we also plot two arrows indicating the displacement of the data points by imposing a stellar correction of 20\% 
(red one) and 10\% (orange one) for the \civalone\ emission \citep[based on][]{Saxena_2022}}.
 }
 \label{fig:BPT}
\end{figure*}

In Fig. \ref{fig:BPT} we present the four different diagrams, which for simplicity we refer to UV-BPT1, UV-BPT2, UV-BPT3 and UV-BPT4 from here on. These diagrams give the advantage that objects can be tracked and diagnosed even if only one of the three lines is detected - in fact, if the S/N of a line is $<2$ we consider 2$|\sigma|$ as a limit. Whenever both lines considered in a certain line ratio are not detected but are both limits, we do not plot that object in the relevant UV-BPT diagram. This means that only UV-BPT3 and UV-BPT4 contain all the 43 objects in our sample.

In each diagram, the dots with error bars and/or limits indicate the position of the sources in our sample; the green triangles represent the position of the X-ray AGNs which have been excluded from our catalog and the green and magenta dots are the predictions from the models described in the previous subsections, respectively from the AGN models \citep{Feltre_2016}, and the SF models \citep{Xiao_2018, Gutkin_2016}. 
It can be seen that the models occupy regions that are quite well although not completely separated in each of the UV-BPT diagrams. In the UV-BPT1 and UV-BPT4 diagrams we also plot the separation between AGN and SF galaxies proposed by \cite{Nakajima_2018} using the distributions of galaxies and AGNs observed in the VUDS survey.

Based on the models considered and on the position of the X-ray detected AGN, we proposed a slightly revised version of the separation between AGN and SF galaxies in the UV-BPT1 diagram which can be written as:
\begin{equation}
 \log(\textrm{C}\textsc{iv}/\textrm{C}\textsc{iii}]) < 4 \log((\textrm{C}\textsc{iv}+\textrm{C}\textsc{iii}])/\textrm{He}\textsc{ii}) - 2;
\end{equation}
For the other UV-BPT diagrams we consider the following separation lines: for UV-BPT2 
\begin{equation}\begin{split}
 \log(\textrm{C}\textsc{iii}]/\textrm{He}\textsc{ii}) & < 0.8 \log(\textrm{C}\textsc{iv}/\textrm{He}\textsc{ii}) + 0.5 \ [\log(\textrm{C}\textsc{iv}/\textrm{He}\textsc{ii})<0.4]\\
 & < 1.5 \ [\log(\textrm{C}\textsc{iv}/\textrm{He}\textsc{ii})\geq0.4].\end{split}
\end{equation}

For UV-BPT3 \citep[not considered by][]{Nakajima_2018}:
\begin{equation}
 \log(\textrm{C}\textsc{iv}/\textrm{He}\textsc{ii}) < \log(\textrm{C}\textsc{iv}/\textrm{C}\textsc{iii}]) + 0.1.
\end{equation}
For UV-BPT4 \citep[same as in][]{Nakajima_2018},
\begin{equation}
\begin{split}
 \textrm{C}\textsc{iv} EW & < 3 \times (\textrm{C}\textsc{iv}/\textrm{He}\textsc{ii}) \ [\textrm{C}\textsc{iv}/\textrm{He}\textsc{ii} <4]\\
&< 12 \ [\textrm{C}\textsc{iv}/\textrm{He}\textsc{ii} \geq 4].
\end{split}
\end{equation}

Not all of 43 sources are in all the diagrams but, when present, their position is always consistently in the AGN or SF part of the diagrams for all cases apart from UV-BPT4. 
In fact we note that the results from UV-BPT4 are more ambiguous since most sources, including some of the X-ray detected NLAGNs are located in an intermediate region in between the AGN and SF models. In addition, it seems that none of the SF models, can produce \textrm{C}\textsc{iv} EW as large as those observed in our sample. This problem was already highlighted in \cite{Saxena_2020}, where we noted that although the low-metallicity binary star models are able to reproduce the UV line ratios of their \textrm{He}\textsc{ii} emitters, they under-predict the \textrm{He}\textsc{ii} EWs. The same applies to the \textrm{C}\textsc{iv} emission which is under-predicted by \textsc{bpass}, as also discussed in \cite{Nakajima_2018, Lefevre_2019}.
For the separation between AGN and SF galaxies we therefore base our classification on the bases of the other 3 UV-BPT diagrams. Specifically we find: 
\begin{itemize}
 \item 4 sources whose position is consistently in the AGN regions in all UV-BPT diagrams: we call these sources "certain AGN" and indicate them with blue dots;
 \item 10 sources whose position is within the AGN regions, but given that the \textrm{C}\textsc{iii} and/or \textrm{He}\textsc{ii} emission lines are actually limits, could move into the SF regions. We call these sources "possible AGN" and indicate them in magenta;
 \item 29 sources whose position is consistently in the SF region for all UV-BPT diagram. We call of course these sources "SF" and indicate them in black.
\end{itemize}
We note that of the 8 X-ray detected LAGNs, 1 source is consistently located within the SF region, while the other are in the AGN region.
This could be due either to a limitation of the models in reproducing all AGN properties or to the fact that in many objects the UV emission line might be of mixed origin.

\subsection{Spectral stacking \label{sec:stack}}
To further shed light on the real nature of the 10 "possible AGN" we apply a spectra stacking technique to infer the average properties of these sources.

\begin{figure*}
 \centering
 \includegraphics[trim=0 140 0 0, width=1\textwidth]{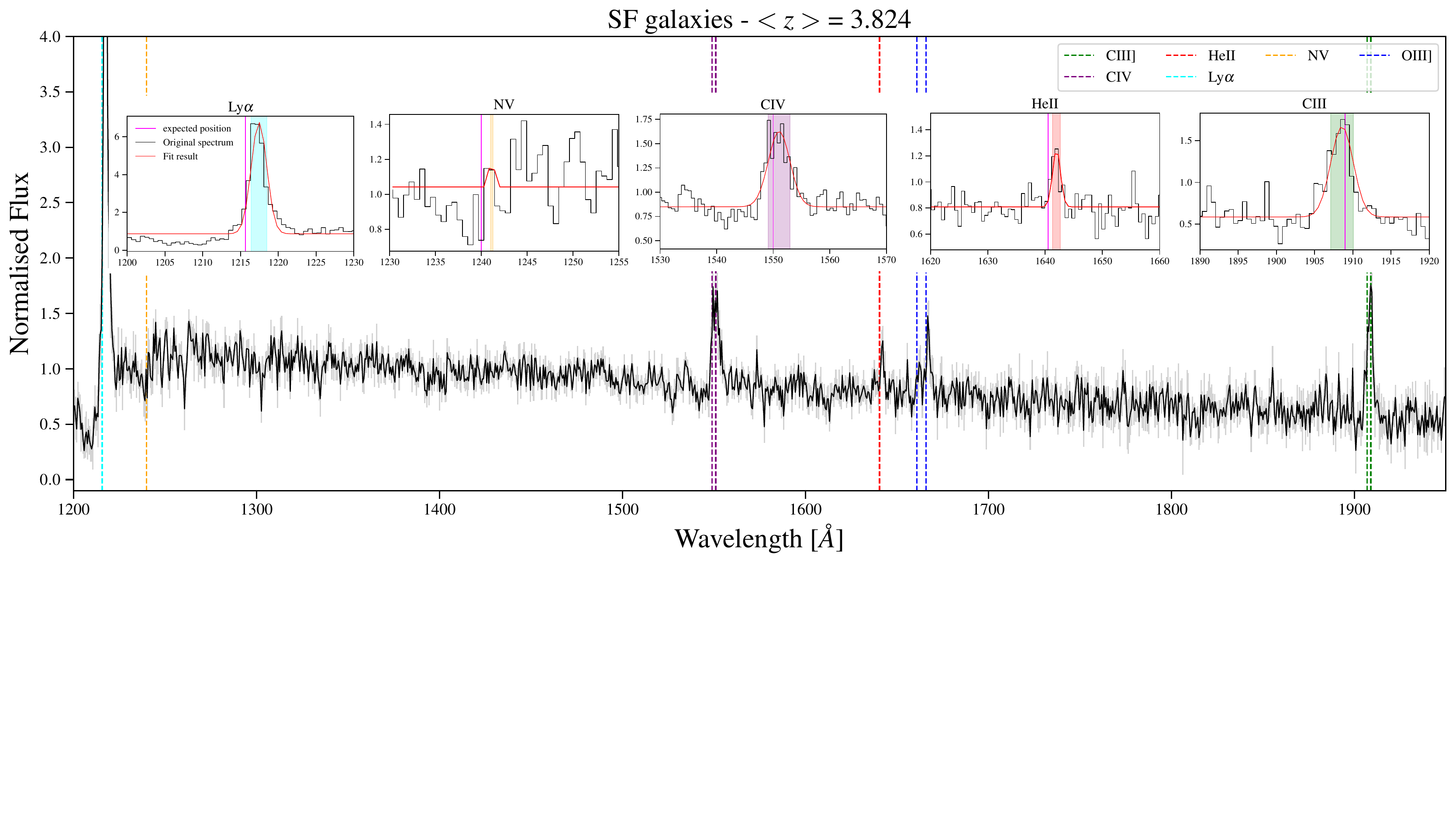}
 \includegraphics[trim=0 140 0 0, width=1\textwidth]{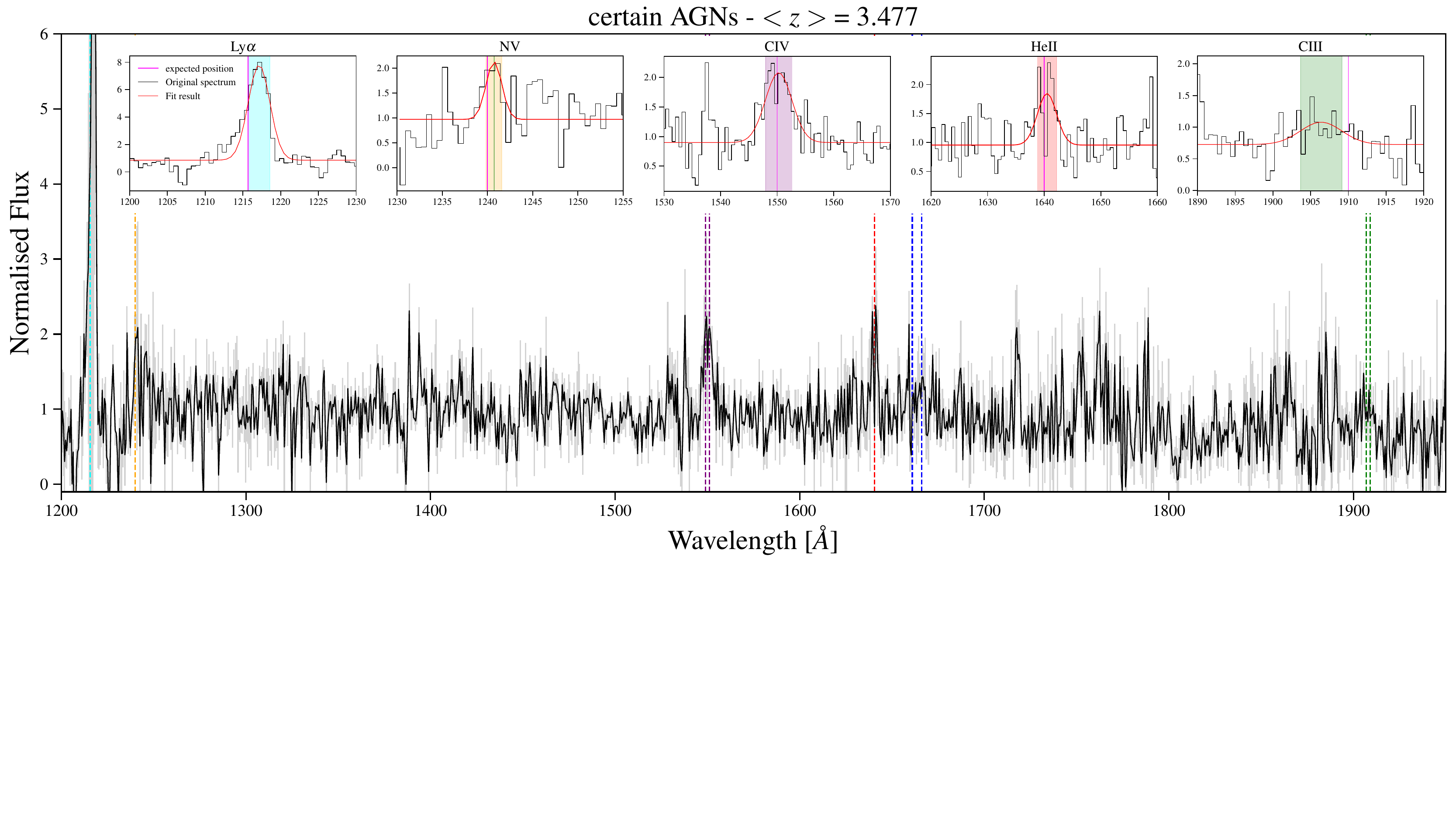}
 \includegraphics[trim=0 180 0 0, width=1\textwidth]{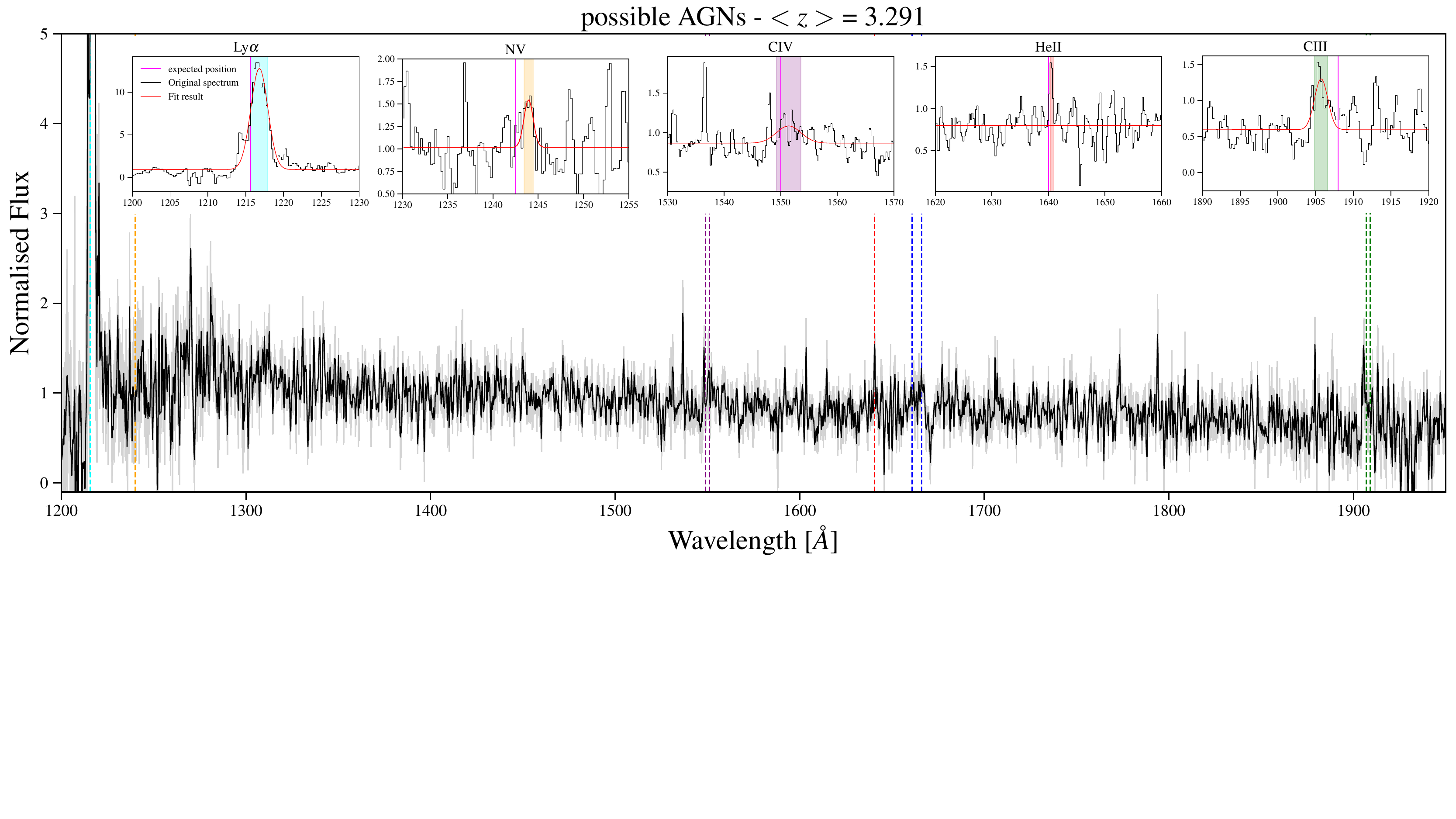}
 \caption{Upper panel. Stacked spectrum (black) of all the 29 SF galaxies of the sample. Middle panel. Stacked spectrum of all the certain 4 AGNs of the sample (no X-ray detection). Lower panel. Stacked spectrum of all the possible 10 AGNs of the sample. The noise obtained from bootstrapping is shown in grey. For clarity, the Gaussian fits with 1 $\sigma$ regions for the emission lines Ly$\alpha$, \textrm{N}\textsc{v}, \textrm{C}\textsc{iv}, \textrm{He}\textsc{ii} and \textrm{C}\textsc{iii} in each stacked spectrum considered in this work are also present. Certain and possible AGN composite show clear broad nebular emission line features in their spectra, while the SF stack show much narrower lines.}
 \label{fig:stack}
 \vspace{-22pt}
\end{figure*}

Following \cite{Saxena_2020}, the stacking is performed by first converting each spectrum to the rest-frame, using the spectroscopic redshift given by the DR4 catalogue.  \textcolor{black}{We point out that the VANDELS redshifts are not the systemic ones, since they are derived by a match to spectral templates
and they are driven by absorption and/or emission lines depending on their strength \citep{Garilli_2021}.
For this reason in the stacks the various lines might not appear at the position expected, e.g. the \textrm{C}\textsc{iii} is not at the exact redshift as we would expect if we had used the systemic redshift for each source.
For our analysis of the stacks we neglect this effect}. The rest-frame spectra are first normalised using the mean flux density value in the rest-frame wavelength range 1300 \AA $\ <\lambda <1500$ \AA. The spectra are then re-sampled to a wavelength grid ranging from 1200 to 2000 \AA, with a step size of 0.582 \AA (the wavelength resolution obtained at a redshift of 3.6, which is the median redshift of VANDELS sources in this work). The errors on the stacked spectra are calculated using bootstrapping: we randomly sample and stack the same number of galaxies from the sample 500 times, and the dispersion on fluxes thus obtained gives the errors (Fig. \ref{fig:stack}).

\begin{figure}
 \centering
 \includegraphics[width=0.45\textwidth]{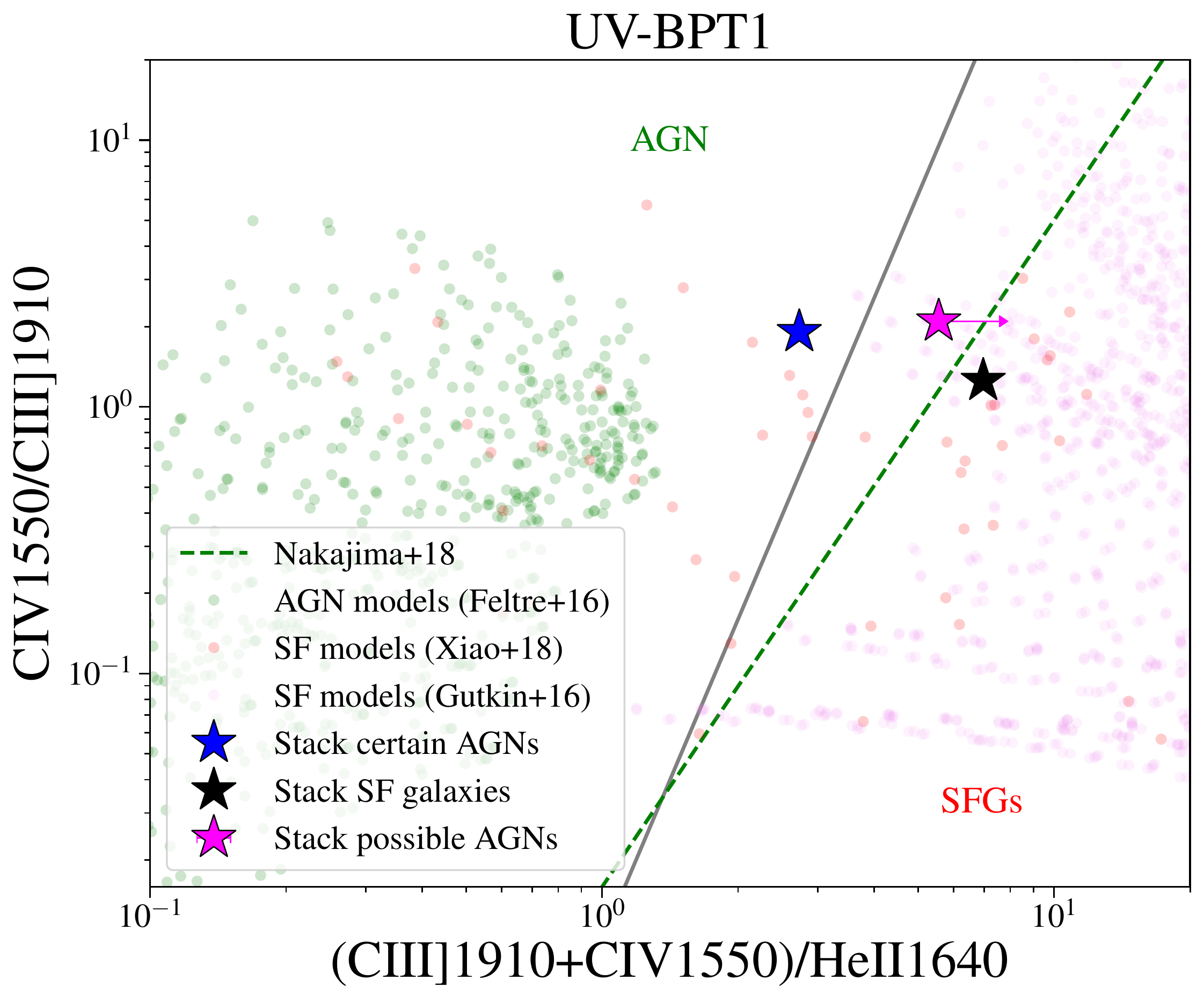}
 \caption{UV diagnostic diagram presenting photoionization models of AGNs \citep{Feltre_2016} and SF galaxies \citep{Xiao_2018, Gutkin_2016}. The separating lines consistently divide the two populations and thus the results from the stack spectra for the SF galaxies, the certain AGNs and the possible AGNs (respectively the stars in black, magenta and blue).}
 \label{fig:BPT_stacks}
\end{figure}

We produce 3 different stacks, one of the possible AGN (10 objects), one with the certain AGN (4 sources) and one with the purely SF galaxies (29 sources). The spectra obviously reveal a wide range of emission and absorption features, with the Ly$\alpha$ emission line standing out in each of the sub-samples. 
The main UV emission lines are fitted with a Gaussian to determine their fluxes and the \textrm{C}\textsc{iv} EW, in order to place the stacks in the various UV-BPT diagrams, an the results are presented in Fig. \ref{fig:BPT_stacks}. We show just one UV-BPT them since the results are consistent for all of them: the blue star represents the sure AGNs, the magenta one the possible AGNs, and the black one the SF galaxies (in accordance with the colors used in Fig. \ref{fig:BPT}). As in the previous UV-BPTs, the green dots are obtained from the AGN models from \cite{Feltre_2016} and the red dots from the SF models from \cite{Xiao_2018, Gutkin_2016}, while the grey lines are the division derived in this work and the green dashed lines refer to \cite{Nakajima_2018}. We note again that the AGN stack is not well reproduced by the AGN models but rather sits in an intermediate position.

 \textcolor{black}{In the SF stacked spectrum (Fig. \ref{fig:stack}) we note that the \textrm{He}\textsc{ii} emission is basically unresolved (FWHM <600 km $\text{s}^{-1}$). This confirms the nebular nature of the \textrm{He}\textsc{ii}: in fact stellar \textrm{He}\textsc{ii} is expected to be much broader of the order of 1500-2000 km $\text{s}^{-1}$
as seen for example in the stacks of the general LBG population \citep{Shapley_2003}. For example \cite{Cassata_2013} used a threshold of 1200 km $\text{s}^{-1}$ to identify narrow \textrm{He}\textsc{ii} emitters of nebular origin. Thus, we can exclude a stellar origin for our sources.}

\subsection{Nature of the "possibile AGNs"}

The properties of the stacked spectra are as expected in the sense that the magenta star is at an intermediate position between the other two: however in all diagrams it is more consistent with the SF regions, indicating that the 10 sources are dominated by SF galaxies rather than AGN.
 
Another check on the nature of the "possible AGN" sample can be obtained by the \textrm{N}\textsc{v} emission. 
For AGNs, as predicted from theoretical and observational studies \citep[e.g.,][]{Hamann_1992, Hamann_1993, Hamann_2002}, we expect a bright \textrm{N}\textsc{v} line. Therefore, we determine the Ly$\alpha$/\textrm{N}\textsc{v} emission line ratio for our three different stacks. For the SF galaxies stacked spectrum the \textrm{N}\textsc{v} line is not detected at all and we can place a limit Ly$\alpha$/\textrm{N}\textsc{v} $> 30$. The stack of the possible AGN presents a Ly$\alpha$/\textrm{N}\textsc{v} = $8.7 \pm 0.6$ that is considerably higher than the certain AGN one (Ly$\alpha$/\textrm{N}\textsc{v} = $2.1 \pm 0.6$), although we have a 2$\sigma$ detection of the \textrm{N}\textsc{v}.  \textcolor{black}{This value is higher than the ratios measured in narrow line AGNs \citep[\lya/\textrm{N}\textsc{v} $\sim 1-2$,][]{Tilvi2016, Hu2017, Sobral2017a}.}

We finally exploit the Chandra X-ray observations: although we have excluded the individually detected X-ray sources since the very beginning, the \cite{Luo_2017} sources are those detected at 5$\sigma$, therefore we cannot exclude faint emission from the "possible AGN" sample. We therefore stacked the X-ray data to check if we detect any flux from the combination of the sources. Unfortunately we can do this in a meaningful way only for the 6 CDFS sources which have much deeper X-ray observations \citep[7 Msec,][]{Luo_2017} while for the UDS field we only have much shallower observations \citep{Kocevski2018}. To stack the X-ray data we follow the same procedure described in \cite{Saxena_2021}. The stack does not produce any detection, although the limit is not very stringent since there are only 6 sources of which one lies at the border of the Chandra field.

In conclusion, from the position of the stack in the four UV-BPT diagrams, the high Ly$\alpha$/\textrm{N}\textsc{v} ratio and the absence of X-ray emission in the stack, we conclude that the sample of 10 possible AGN must be dominated by SF galaxies, although we cannot exclude a (faint) AGN contribution.
In the following we include these 10 sources in the SF sample, bringing it to 39 galaxies in total, to study their physical properties.

\section{Physical properties of the SF \textrm{C}\textsc{iv} emitters}\label{sec5}

The combined synergy of both space-based and ground-based telescopes in the VANDELS fields provide an excellent multi-wavelength dataset from ultraviolet (UV) to mid-infrared (MIR), allowing accurate determination of the physical integrated properties of the VANDELS galaxies. We now describe the SED fitting tool that we have used to determine the physical properties, the procedure adopted and the main results.

\subsection{The \textsc{BEAGLE} tool}\label{sec:beagle}

The code used for the spectral energy distribution (SED) fitting is the \textsc{BEAGLE} (BayEsian Analysis of GaLaxy sEds) software \citep{Chevallard_2016} which was chosen for its versatility: through the Bruzual \& Charlot libraries it allows us to model in a coherent way the continuous radiation coming from the stars of the galaxies and the nebular component which is modeled with \textsc{Cloudy}. The code also includes several prescriptions to describe the dust attenuation and the radiation transfer through the intergalactic medium. Stellar formation and chemical enrichment histories are included both with parametric and non-parametric prescription. Finally, \textsc{BEAGLE} can simultaneously fit the photometry and the spectra (or individual spectral indices) of the galaxies. \textsc{BEAGLE} was run on all the VANDELS parent sample using v0.24.5 which employs the most recent version of the \cite{Bruzual_2003} stellar population synthesis models \citep[see][for details]{Vidal_2017}. The full description of the method and the results on the entire sample will be discussed in a companion paper (Castellano et al. in prep.). We provide here only a brief description of the most important parameters adopted: in particular we assume a delayed star formation history (SFH), which is the most flexible parametric SFH used by the code, with a current SF timescale (i.e. duration of the current episode of star formation) of 10 Million years.  \textcolor{black}{The escape fractions $f_{esc}$ assumed in the BEAGLE run are set to 0 for all galaxies, regardless of the presence of the \civalone\ emission.}
We used a \cite{Chabrier_2003} Initial Mass Function, metallicity in the range $-2.2 \leq \log(Z/Z_\sun) \leq 0.25$ and we treated the attenuation following the \cite{Charlot_2000} model combined with the \cite{Chevallard_2013} prescriptions for geometry and inclination effects.

For each source, \textsc{BEAGLE} is run on the combination of photometry, coming from the official 13 band VANDELS catalog presented in \cite{Garilli_2021}, and spectral indices i.e., the fluxes of the \textrm{He}\textsc{ii}, \textrm{O}\textsc{iii} and \textrm{C}\textsc{iii} emission lines as well as the precise spectral windows used (respectively [1634-1654], [1663-1668] and [1897-1919]). After some tests we decided not to provide the measured \textrm{C}\textsc{iv} emission to \textsc{BEAGLE} as a further spectral index to fit, both to have an unbiased estimate of the physical parameters, and also because as just discussed in Sect. \ref{sec:AGN_SF}, there is a well known inability of stellar population models, even when including binary stars, to reproduce some UV line fluxes \citep[also see][]{Saxena_2022}.
 \textcolor{black}{We also  tested how the choice of current SF timescale could change our results. In particular  this is relevant since the EWs of the emission lines are strongly dependant on the current SF.  As a result even when reducing the SF timescale to 5 Myrs, the physical parameters remain very stable with the exception of $\xi_{ion}$, which becomes slightly higher ($\sim$ 0.1 dex) for all galaxies. Note that the inferred $\xi_{ion}$ from \textsc{BEAGLE} is almost independent from the initially assumed $f_{esc}$: this value is determined by the intrinsic ionizing SED, which is already adjusted for relatively low metallicities and young ages. The change in $\xi_{ion}$ is substantial only for very high $f_{esc}$ \citep[see e.g.,][]{Marques-Chaves2022}.}
The official VANDELS redshift was assumed in all cases for the SED fitting.

As a result, \textsc{BEAGLE} provides the best SED fit and the best physical parameters, with relative uncertainties: specifically total stellar mass, star formation rate, dust attenuation, age and $\xi_{ion}$ were retrieved. For a complete list of the selected physical parameters and the range provided \textit{a priori} to fit all the sources of the sample, see Table \ref{tab:BEAGLE} in the Appendix.


\subsection{Results and comparison to parent population}\label{sec5.2}
\begin{figure}
 \centering
 \includegraphics[width=0.5\textwidth]{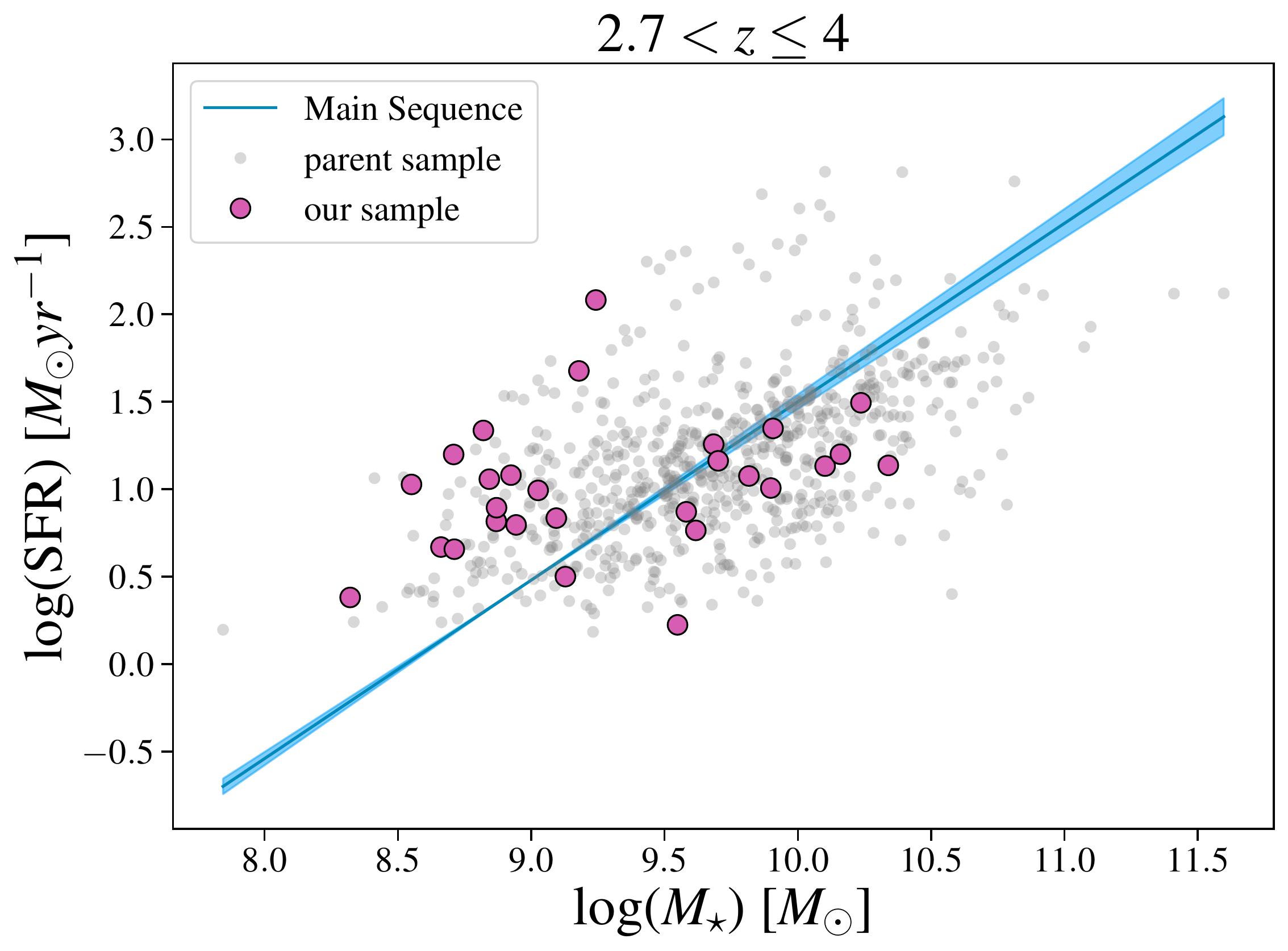}\quad
 \includegraphics[width=0.5\textwidth]{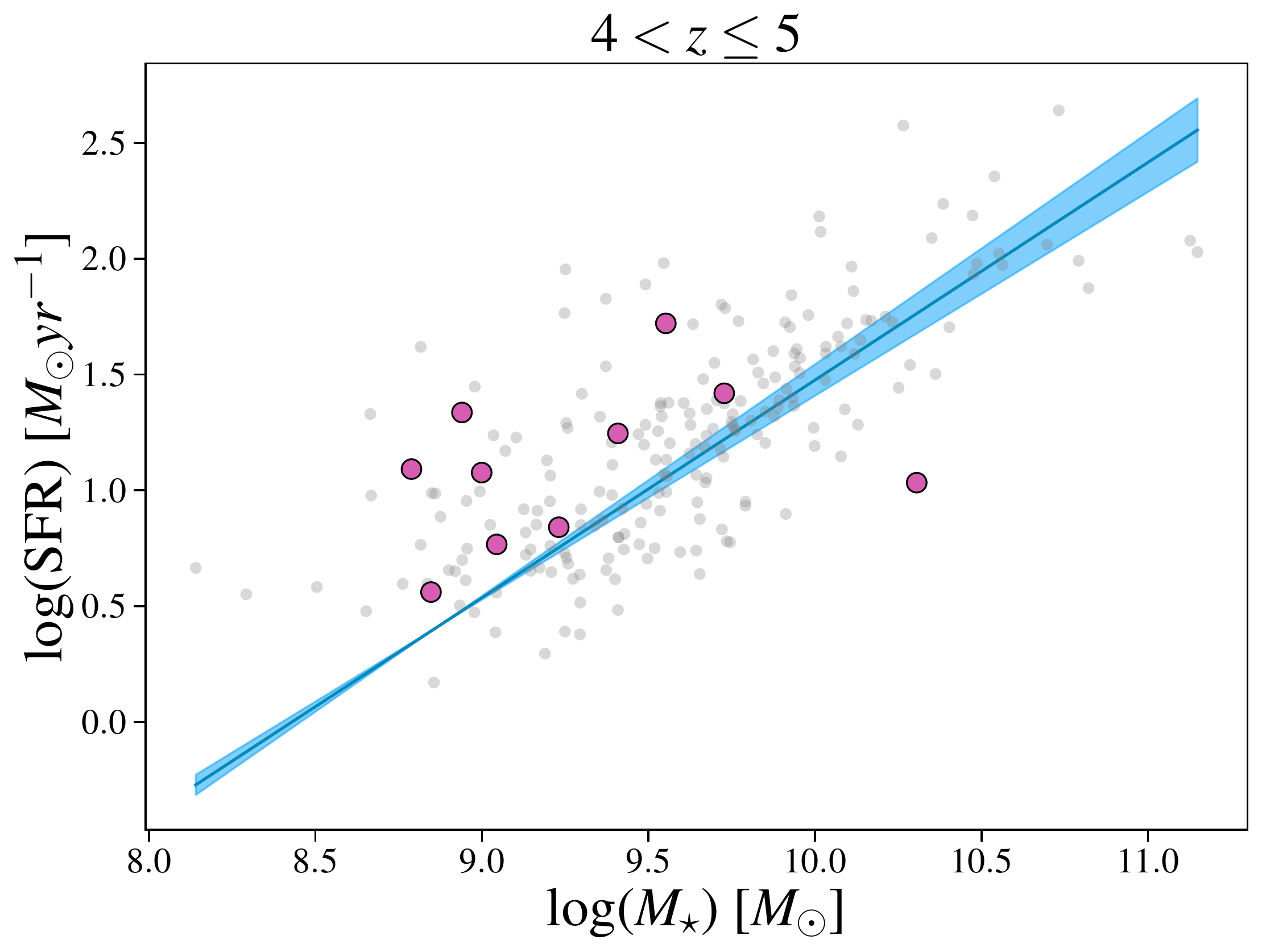}
 \caption{Stellar mass vs SFR relation for our \textrm{C}\textsc{iv} emitters (purple dots) and the parent sample (grey dots). The blue line shows the main sequence respectively at $3<z\leq 4$, $4<z\leq5$ from \cite{Santini2017}.}
 \label{fig:main_seq}
\end{figure}

Using the derived physical properties we searched for systematic differences between the SF sample of the \textrm{C}\textsc{iv} emitters (39 galaxies) and the parent sample. From the parent sample we have removed all X-ray emitters and all possible AGN as defined in Bongiorno et al. (in prep.). 
In Fig. \ref{fig:main_seq} we present the location of our 
\textrm{C}\textsc{iv} emitters on the main sequence i.e. the stellar mass vs SFR plot: given that our sample spans a rather large redshift range, we split the plot into two redshift bins, $2.7<z\leq4$, and $4<z\leq5$.
The blue lines are obtained using the main sequence as a function of redshift described in \citet{Santini2017}. 
The VANDELS sources as well as the \textrm{C}\textsc{iv} emitters sub-sample are located more or less on the main sequence of star forming galaxies, with perhaps a small bias in the high redshift bin, where the (few) objects are mostly located on or above the MS.

\begin{figure}
 \centering
 \subfloat{\includegraphics[width=0.4\textwidth]{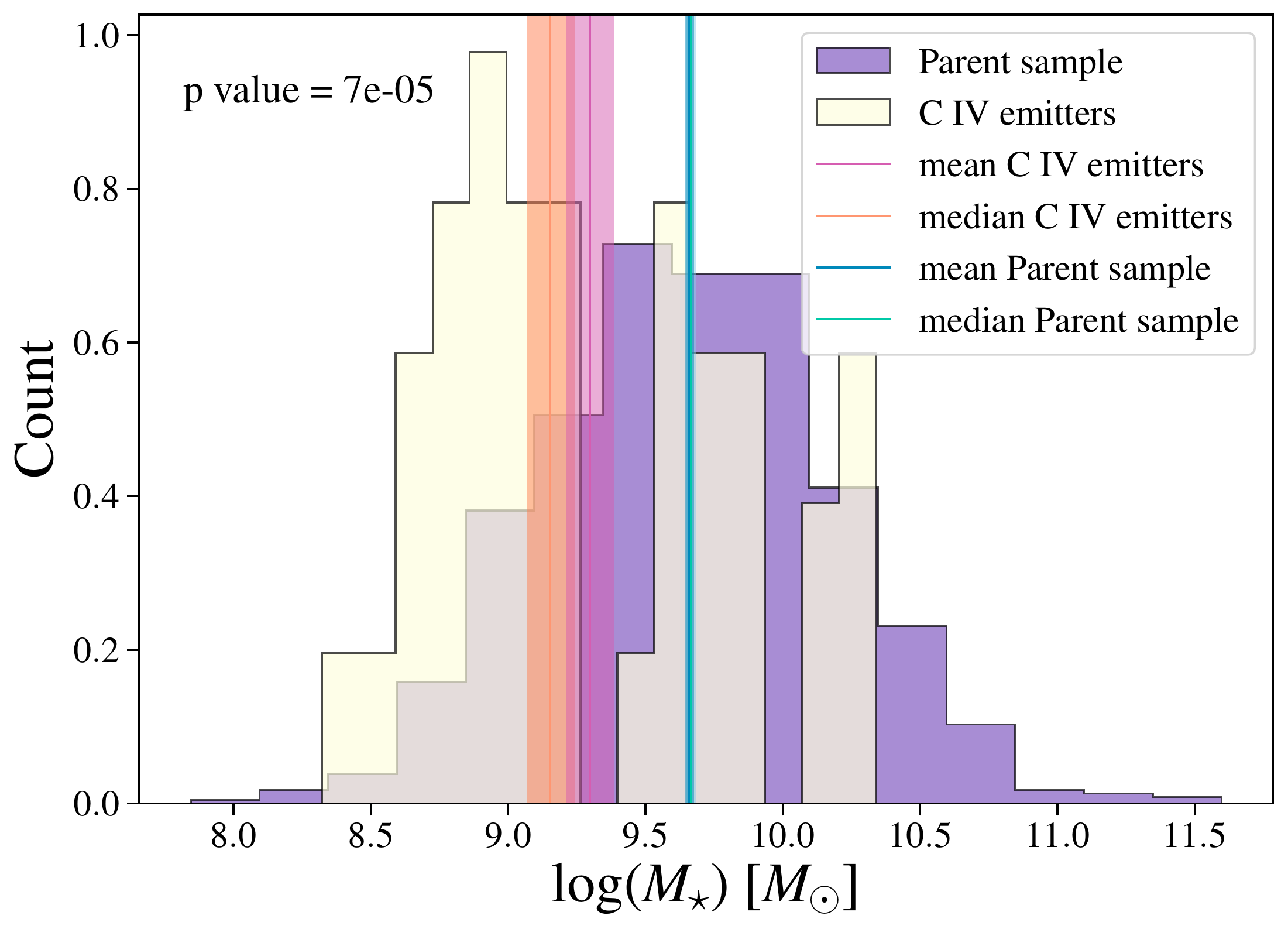}} \quad
 \subfloat{\includegraphics[width=0.4\textwidth]{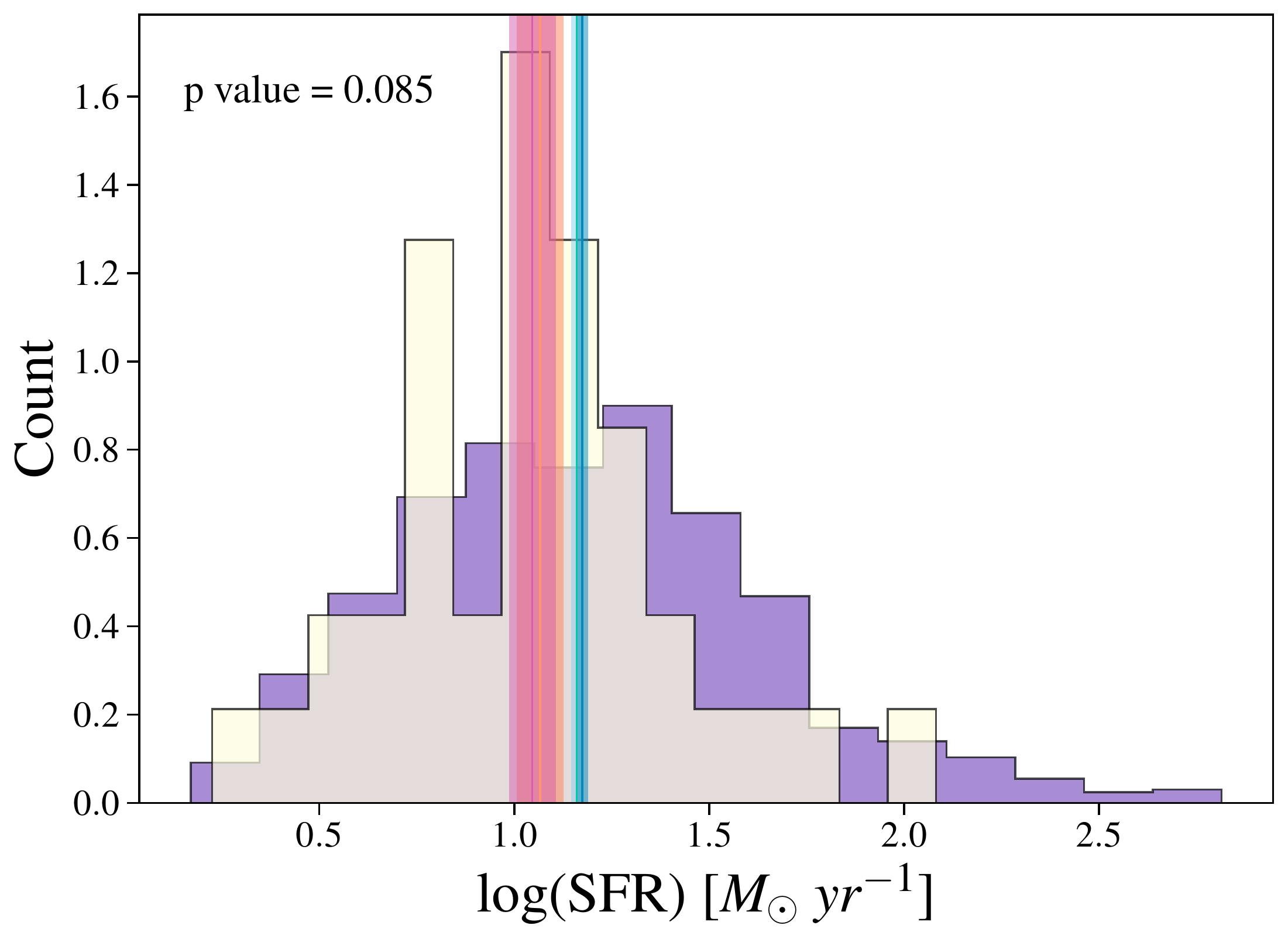}}\quad
 \subfloat{\includegraphics[width=0.4\textwidth]{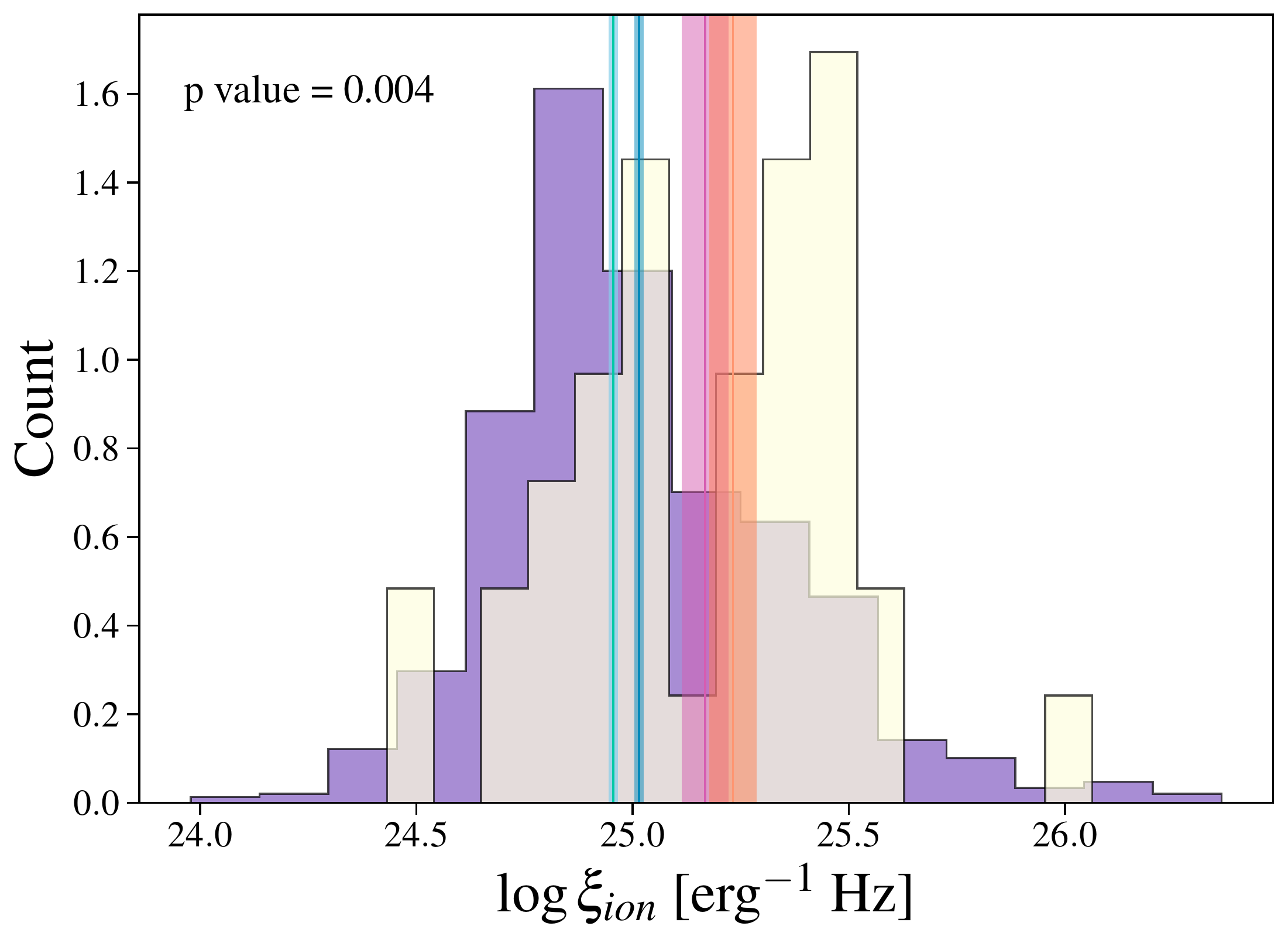}}\quad
 \subfloat{\includegraphics[width=0.4\textwidth]{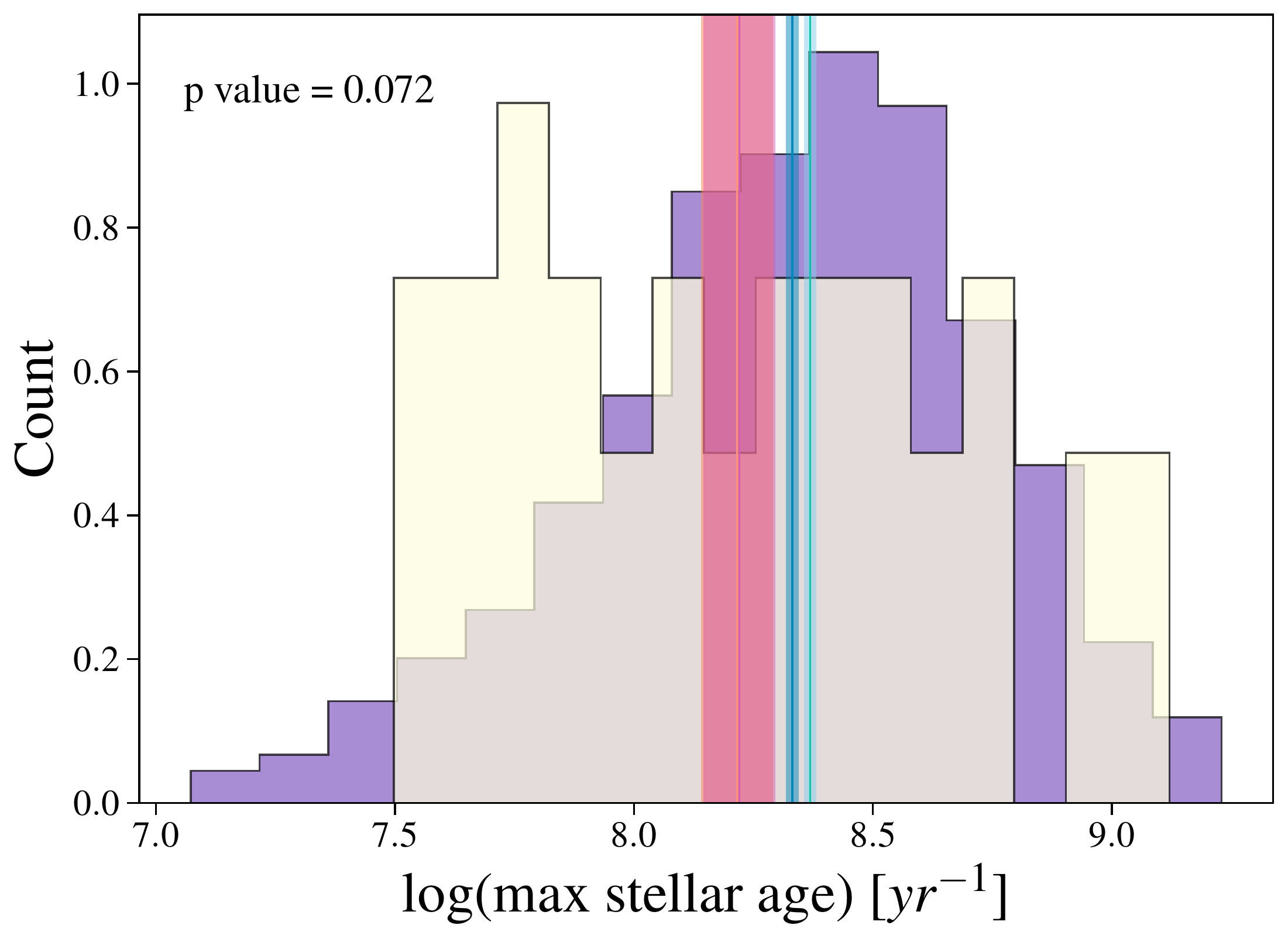}}
 \caption{Derived stellar mass, SFR, ionizing photon production efficiency $\xi_{ion}$ and stellar age of the sources identified as star forming, determined by \textsc{BEAGLE} (light distribution). For comparison, those of the total sample in the same redshift range are also shown in the Figure (purple distribution). The figure also shows the mean and median with their standard deviations for the distributions and the p-value from the KS test for the main physical properties of the analyzed sample.}
 \label{fig:comparison}
 \vspace{-15pt}
\end{figure}

In Fig. \ref{fig:comparison} we then show the normalized distributions of the derived stellar masses, the SFRs, the photon production efficiencies $\xi_{ion}$ and the stellar ages of the \textrm{C}\textsc{iv} galaxies and the parent population. In each panel we also indicate the mean and median values for the two samples with vertical regions, whose widths represent the uncertainty.
From the panels we can see that the \textrm{C}\textsc{iv} emitters tend to have considerable lower masses, slightly lower SFR, higher $\xi_{ion}$ and slightly lower stellar ages compared to the parent sample. 
To assess the significance of these differences, we employ the common Kolmogorov-Smirnov (KS) test.
From the p-values (reported in the figures) we conclude that the mass and $\xi_{ion}$ of the \textrm{C}\textsc{iv} emitters and the parent sample are statistically different, while the max\_stellar\_age and the SFR are compatible with being randomly drawn.
In Castellano et al. (in prep.) we find a strong link between the sSFR and  $\xi_{ion}$: indeed if we look at our \textrm{C}\textsc{iv} emitters they have slightly higher sSFR compared to the parent population, which could explain the higher $\xi_{ion}$. This link and its implications will further discussed in that paper. 
 \textcolor{black}{Even using a SF timescale of 5 Myrs, with all values of  $\xi_{ion}$ being slightly higher, the difference between \civalone\ emitters and the parent sample is still significant.}

In conclusion our \textrm{C}\textsc{iv} emitters are in general low mass objects which have an higher photon production rate compared to the general star forming population at the same redshift.

\subsection{Average stellar metallicity}
The link between the presence of nebular \textrm{C}\textsc{iv} emission and low metallicity stellar population has been noted both at high and low redshift. For example \cite{Senchyna_2017} showed that \textrm{C}\textsc{iv} emission may be ubiquitous in extremely metal-poor galaxies with very high specific star formation rates, although it reaches large EWs only if the ISM is exposed to the hard stellar ionizing spectra produced at extremely low metallicities (12 + log O/H $\lesssim$ 7.7). 

To derive the stellar metallicity from our spectra, we use the method developed by \cite{calabro}, which is based on stellar photospheric absorption features at 1501 \AA\ and 1719 \AA, calibrated with Starburst99 models and are largely unaffected by stellar age, dust, IMF, nebular continuum, or interstellar absorption. The method can be applied to spectra with a S/N of 15-20 in the continuum to avoid large uncertainty, and therefore we cannot derive metallicities for individual sources but only average values from stacked spectra. We produced a stack of the 39 SF galaxies following the procedure outlined in the previous section. The metallicity we derive is $Z \sim 0.26 Z_\odot$ for the sample of 29 purely SF galaxies and a slightly lower (but consistent within the uncertainties) value of $Z \sim 0.1 Z_\odot$ for the combined sample of 39 galaxies. 
Looking at the stellar mass-stellar metallicity relation by \cite{calabro} we can see that these low metallicities (about $0.15 Z_\odot$) are perfectly consistent with the general trend for $z\sim$ 3 galaxies with similar masses, i.e. the \textrm{C}\textsc{iv} emitting galaxies do not show lower 
metallicities compared to their parent sample, as also found by \cite{Saxena_2022}. This is therefore at odds with the low redshift universe where \textrm{C}\textsc{iv} emitters have almost always an inferred stellar (and gas phase) metallicities are extremely low \citep[e.g.,][]{Berg_2019, Senchyna_2021}. However in terms of "absolute" metallicity, $z\sim$3 \textrm{C}\textsc{iv} emitters have a similar metallicity as the local universe emitters: this is of course due to the fact that at $z \sim 3.5$ the average metallicity of star forming galaxies is a factor of 0.6 dex lower than in the local universe \citep{Cullen_2020, calabro}. 

\begin{figure}
 \centering
 \includegraphics[width=0.5\textwidth]{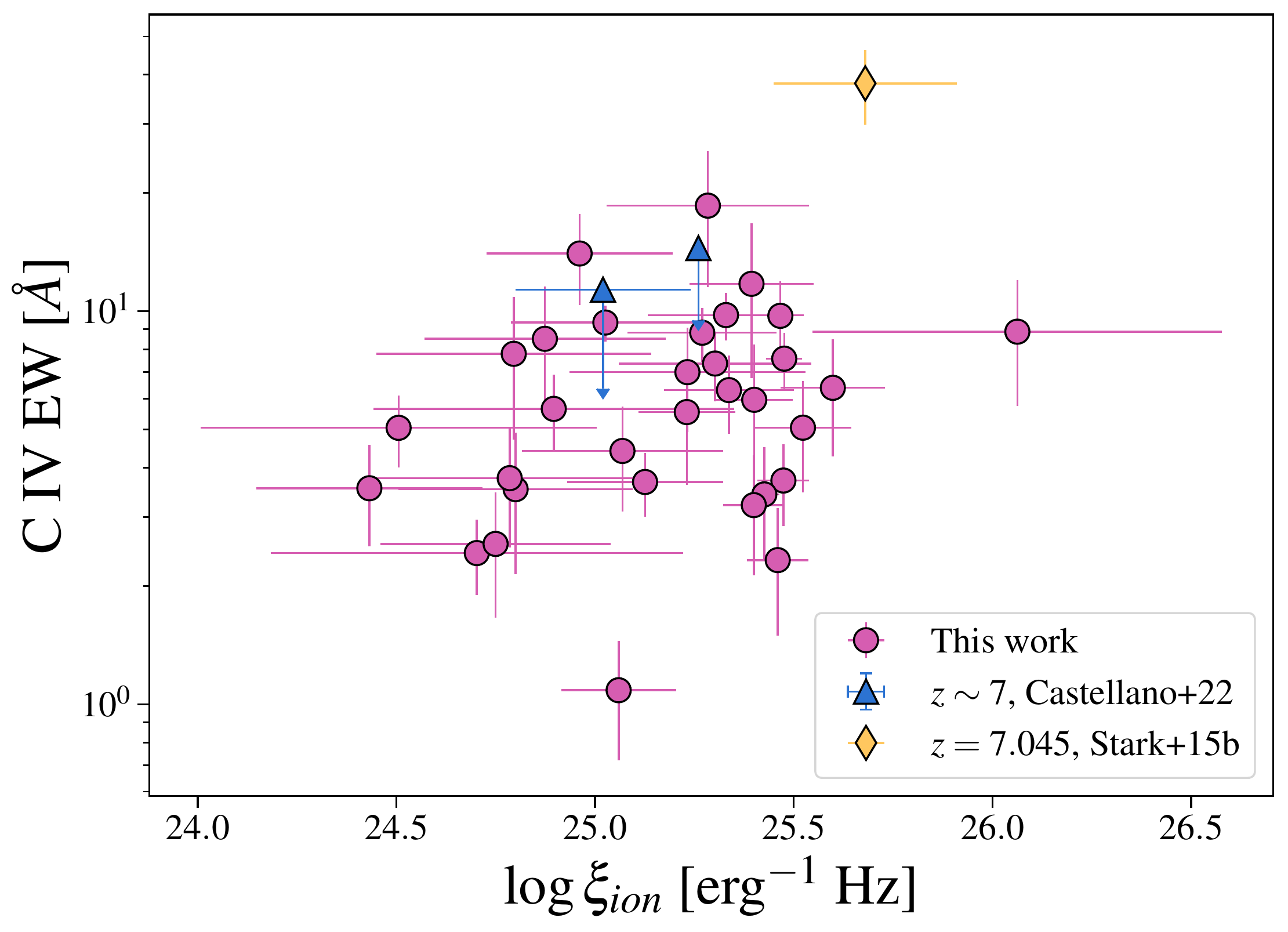}
 \caption{Distribution of the measured EW(\textrm{C}\textsc{iv}) and the $\xi_{ion}$ for the selected star forming galaxies identified in both CDFS and UDS fields (purple points) compared with $z\sim 7$ objects from literature: \cite{Castellano_2022} (blue points) and \cite{Stark_2015} (orange point).}
 \label{fig:civ_vs_xiion}
\end{figure}

\section{Discussion: the ionizing properties of \textrm{C}\textsc{iv} emitters}\label{sec:LyC_leakers}

In the previous sections we have found that the VANDELS \textrm{C}\textsc{iv} emitters are in general galaxies with low masses and high photon production efficiency compared to their parent population. They have stellar metallicities of $0.1-0.15 Z_\sun$ consistent with the stellar-mass stellar metallicity relation at redshift 3.5. Most of our \textrm{C}\textsc{iv} emitters show also a prominent Ly$\alpha$ emission, and we observe a strong correlation between the \textrm{C}\textsc{iv} and the $Ly\alpha$ flux (shown in Fig. \ref{fig: line_comparison}). 
We now investigate if the \textrm{C}\textsc{iv} emitters have properties that could make them potentially good ionizers.
 
An investigation of ionizing efficiency is the first step in identifying good candidates for cosmic reionizers. Photon production efficiency expectations depend on several factors, including the initial mass function, the SFH, the evolution of individual stars, the stellar metallicity, and potential binary interactions between stars \citep[e.g.,][]{Zackrisson2011, Zackrisson2013, Zackrisson2017, Eldridge_2017, Stanway2018, Stanway2019}, but they agree in identifying the expected value of $\log\xi_{ion}$ around 25.3 to ionize the IGM at $z>6$ \citep{Robertson2013, Robertson2015}. This prediction is confirmed both from the local universe and intermediate and high redshifts' observations \citep[e.g.,][]{Matthee2017a, Izotov2017, Nakajima2018b, Shivaei2018, Lam2019, Bouwens2016, Stark_2015, Stark2017, Atek2022, Castellano_2022}. $\log\xi_{ion}$ is also known to have a higher-than-average value when strong Ly$\alpha$ \citep[e.g.,][]{Harikane2018, Sobral2019, Cullen_2020} and UV-rest frame emission lines \citep[e.g.,][]{Nakajima_2016, Tang_2019} are present. Using the results obtained from our sample, we examine the dependence of the $\log\xi_{ion}$ (calculated, as stated in Sec. \ref{sec5.2}, without considering \textrm{C}\textsc{iv} emission) with respect to the $EW_0$ of the nebular \textrm{C}\textsc{iv}. The results are shown in Fig. \ref{fig:civ_vs_xiion}. We also compare our values with $z\sim 7$ objects from literature \citep{Stark_2015, Castellano_2022} with \textrm{C}\textsc{iv} detection. According to this analysis, high \textrm{C}\textsc{iv} emission would be accompanied by an increase in $\xi_{ion}$ values, which are also within the range expected for cosmic reionization. For a detailed examination of $\xi_{ion}$ in relation to the VANDELS sample proprieties, see Castellano et al. (in prep.).
 
We now turn to the other relevant necessary condition for an ionizer, i.e. the LyC escape fraction. We know that the presence of strong Ly$\alpha$ emission is tightly linked to the escape of LyC photons. Galaxies with moderate values of $f_{esc}$ are in general characterised by strong Ly$\alpha$ emission, as observed both in individually detected galaxies \citep[e.g.,][at high and low redshifts, respectively]{Pahl_2021, Begley2022, Gazagnes_2020, Flury2022b} and via stacked spectroscopic studies \citep[e.g.,][]{Marchi_2018, Steidel_2018A}. Such a tight relation is also predicted by models which show that both ionising continuum flux and Ly$\alpha$ photons can escape through the same ionised channels in the interstellar medium \citep{Rivera-Thorsen_2015, Verhamme_2015, Dijkstra_2016, Rivera-Thorsen_2017, Verhamme2017, Steidel_2018A, Marchi_2018, Izotov_2020, Gazagnes_2020, Izotov_2021}. 
We checked the 37 VANDELS \textrm{C}\textsc{iv} emitters that have their Ly$\alpha$ in the observed spectral range (all those at $z>3$), and only one has Ly$\alpha$ in absorption, while 24 have Ly$\alpha$ EW larger than 20\AA, with several above 100\AA. This would imply that a good fraction of the \textrm{C}\textsc{iv} emitters might also be galaxies with non-negligible ionizing escape fractions. 
As shown by \cite{Verhamme2017}, the Ly$\alpha$ profiles of strong LyC leakers are in general all double-peaked with a small peak separation, in agreement with our theoretical expectation: unfortunately due to the relatively low resolution of the VANDELS spectra we are not able to determine if this is also the case for our \textrm{C}\textsc{iv} emitters. 
 
\begin{figure}
 \centering
 \subfloat{\includegraphics[width=0.5\textwidth]{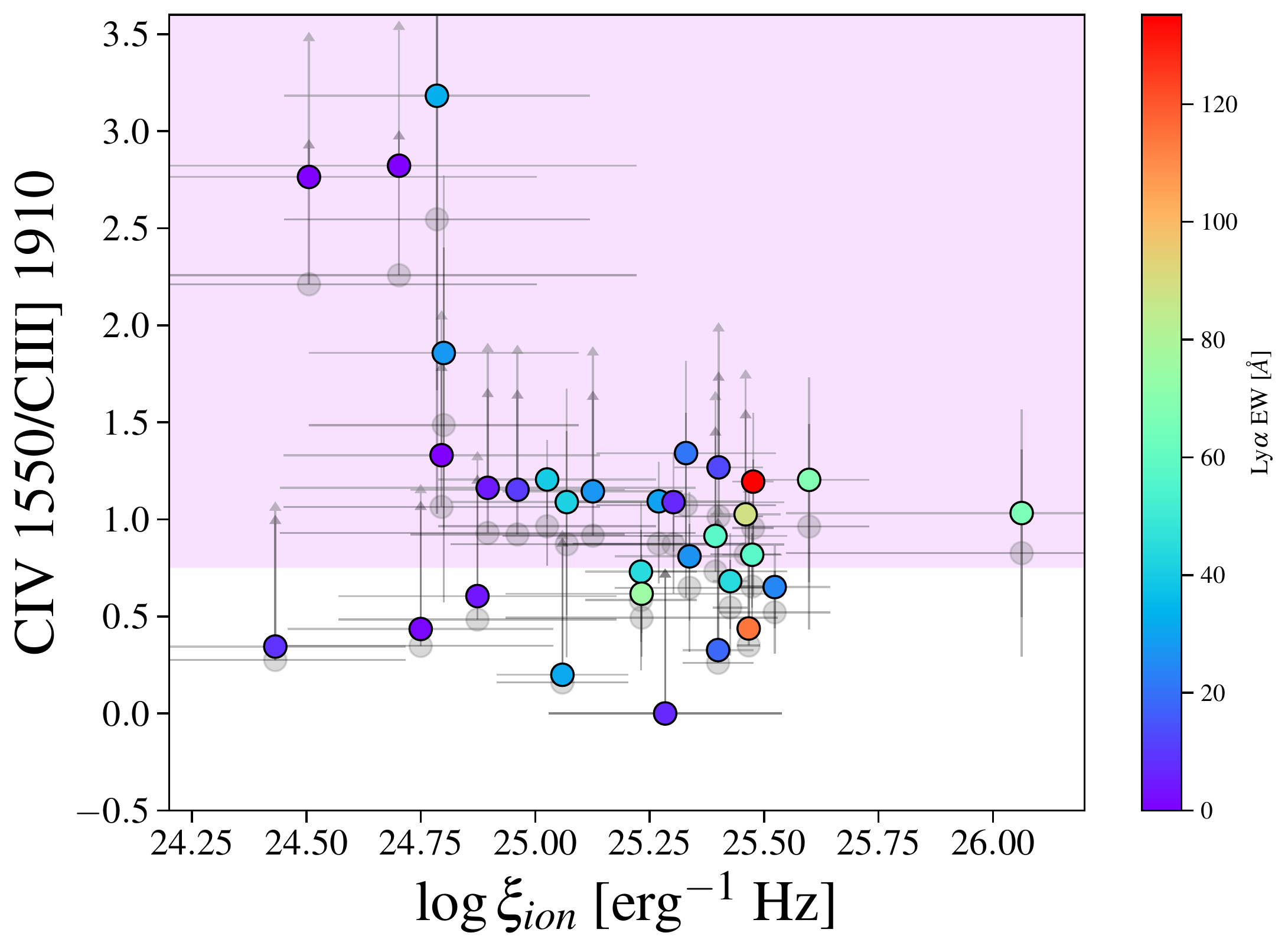}}\quad
 \caption{Ratio of \textrm{C}\textsc{iv} and \textrm{C}\textsc{iii}] versus the $\xi_{ion}$ for individual galaxies in this study. The shaded area limits C34>0.75, that was empirically suggested by \cite{Schaerer_2022} to host strong LyC emitting galaxies ($f_{esc} > 0.1$). The sources are colored using the Ly$\alpha$ EW values. We find that 21 out of 32 galaxies in our sub-sample have \textrm{C}\textsc{iv}/\textrm{C}\textsc{iii}] ratios > 0.75.  \textcolor{black}{In grey we  plot the same population assuming an average  stellar correction of 20\%.}}
 \label{fig:leakers}
\end{figure}

We further checked whether the properties of our \textrm{C}\textsc{iv} emitters are similar to the 
LyC leakers of \citet{Schaerer_2022}. They 
 define \textit{strong leakers} as those galaxies with a LyC escape fraction above 10\% (i.e., $f_{esc}$ > 0.1) and found that these galaxies show not only the presence of \textrm{C}\textsc{iv} emission but also a high ratio, C43 = \textrm{C}\textsc{iv}/\textrm{C}\textsc{iii}] > 0.75. 
In Fig. \ref{fig:leakers} we show the C43 values for our sample, in the cases when the \textrm{C}\textsc{iii}] emission is measurable (in 10 out of 39 cases, the \textrm{C}\textsc{iii}] falls outside the range included in the VANDELS spectra). For 22 galaxies the C43 ratio is clearly $> 0.75$ and only in 4 cases it is smaller, with the rest being galaxies in which the relatively high \textrm{C}\textsc{iii}] limit prevents us from drawing any conclusion. In any case at least two thirds of the \textrm{C}\textsc{iv} emitters seem to have properties consistent with the low redshift LyC leakers presented in \cite{Schaerer_2022}.  \textcolor{black}{Since in \cite{Schaerer_2022} the \civalone\ emission was purely nebular in origin, in this Figure we also show how the results would be effected by considering the possible stellar contribution. The grey symbols in Fig. \ref{fig:leakers} thus represent the effect of assuming  a 20\% stellar contribution correction on the sample. We note that only two sources move outside the previously defined leakers' region, but given the high uncertainties on C43, the basic conclusions remain unchanged.}

In a companion paper \citep{Saxena_2022} we have investigated a sub-sample of the bright VANDELS \textrm{C}\textsc{iv} emitters by means of spectral stacking to understand if they could indeed be good tracers of hydrogen ionising LyC photon leakage using a variety of indicators. We found that equivalent width and offset of the Ly$\alpha$ peak from the systemic redshift of galaxies is indicative of significant LyC leakage. These characteristics are consistent with the predictions of e.g. \cite{Verhamme_2015} and \cite{Dijkstra_2014} of high Ly$\alpha$ and LyC escape in galaxies.
 In addition, the depth and velocity offsets of low-ionisation interstellar absorption features in the stack, are indicative of low column density channels through which LyC photons may escape \citep[e.g.,][]{Chisholm_2020, Saldana-Lopez_2022}, with their low covering fractions suggesting LyC $f_{esc} \simeq 0.05 - 0.30$. 
 
In conclusion, a consistent picture emerges, in this work and in the parallel analysis by \cite{Saxena_2022} in that a substantial fraction of galaxies showing nebular \textrm{C}\textsc{iv} emission, could be galaxies with significant production and escape of LyC photons. They could therefore be the best analogs of those galaxies that mostly contributed to cosmic reionization at $z>7$.  \textcolor{black}{We have shown that \textrm{C}\textsc{iv} emitters do not differ significantly from low-mass galaxies apart from enhanced $\xi_{ion}$ (which is probably insufficient to explain the strong UV emission lines associated with \textrm{C}\textsc{iv}). So if the integrated properties (SFR, metallicity, etc.) are equal, the intense UV emission must be instead linked to a peculiar ISM structure. For example, it has been shown that in order to accurately reproduce such high UV emission lines,  gas-denser regions and lower-density regions must have different ionizations \citep[][]{Berg2021, Schaerer_2022} in order to facilitate the escape of ionizing photons. The \lya\ flux at systemic velocity \citep{Saxena2022a}  observed in the stack of the bright \civalone\ emitters supports this scenario, suggesting that it is the ISM properties that make the \textrm{C}\textsc{iv} emitting galaxies different from the parent population.}

 \textcolor{black}{Finally,} we identify the best candidate ionizers in our sample by  selecting  all galaxies having C43 = \textrm{C}\textsc{iv}/\textrm{C}\textsc{iii}] > 0.75, EW(Ly$\alpha$)>20\AA, and $\xi_{ion} > 25 $ (the median of the parent sample): in total we have 12 sources with these combined properties and we indicate them in red in Table \ref{Tab_prop}. Note that Ly$\alpha$ and \textrm{C}\textsc{iii}] are outside of the observed range respectively for 2 and 10 of the SF sample, so the 12 candidates are selected from a usable sample of only 31 galaxies, thus representing almost 40\% of the galaxies.

\section{Summary and conclusions}\label{sec:conclusions}

In this paper we have presented a detailed investigation of star forming galaxies at $2.5 \leq z \leq 5$ which show significant nebular \textrm{C}\textsc{iv} emission, drawn from the VANDELS survey \citep{McLure_2018, Pentericci_2018}. We summarize our findings as follows: 

\begin{enumerate}
 \item At $2.5 \leq z \leq 5$, \textrm{C}\textsc{iv} emitters represent just $\sim 4\%$ of the VANDELS star forming population (933 sources). Their $EW_0s$ range from 1 to 29 \AA \ (Sect. \ref{sec3}). The \textrm{C}\textsc{iv} strength correlates well with the strength of the other UV emission lines.
 \item Using the UV-BPTs diagrams, we have identified those sources that are powered by stellar photoionization, indicating massive stars as the only ionization source (Sect. \ref{sec4}), opposed to an AGN origin. Out of the 43 
 initially selected \textrm{C}\textsc{iv} emitters, only 4 are consistently found in the AGN region of the diagrams. For 10 sources which could possible be AGN, we resort to spectral stacking analysis and confirm that are most probably SF galaxies, bringing the final sample to 39 sources.
 \item Using \textsc{BEAGLE} \citep{Chevallard_2016}, by fitting photometry and the spectral indices we derived the physical properties of the SF \textrm{C}\textsc{iv} emitters (total stellar mass, SFR, $\xi_{ion}$, stellar age). We also derive their average stellar metallicity following \cite{calabro} method based on weak absorption lines.
 We find that \textrm{C}\textsc{iv} emitters 
  are low-mass galaxies residing on the main sequence. Their metallicity is consistent with the general SF population at the same redshift, and similar to the metallicity of \textrm{C}\textsc{iv} in the local universe. \textrm{C}\textsc{iv} emitters  \textcolor{black}{have on average similar integrated physical properties when compared to galaxies with no \textrm{C}\textsc{iv} emission at $2.5 \leq z \leq 5$, with the exception of a higher $\xi_{ion}$ average value.} 
 \item Besides having high production of ionizing photons, there are several indications that \textrm{C}\textsc{iv} emitters have also high escape fractions of LyC photons: this includes the presence of bright Ly$\alpha$ emission and a high \textrm{C}\textsc{iv}/\textrm{C}\textsc{iii}] ratio which was recently proposed as one of the most promising indicators of a possible LyC leakage by \cite{Schaerer_2022}. In Sect. \ref{sec:LyC_leakers}, we identify 12 galaxies as the best analogs to cosmic reionizers, showing the combination of high $\xi_{ion}> 25$, and EW(Ly$\alpha$)>20\AA \ and \textrm{C}\textsc{iv}/\textrm{C}\textsc{iii}] > 0.75, indicative of high LyC escape fraction.
\end{enumerate}

Amongst the inferred properties, the most uncertain is obviously the escape fraction: in this study we only have presented indirect evidence that the galaxies could have high $f_{esc}$. Unfortunately, the ultra deep VANDELS spectra only start at 4800 \AA, which means that a direct measurement of the Lyman continuum flux is impossible with current data. However, at this redshift it is possible to detect the Lyman continuum flux at $\sim$ 3800-4000 \AA\ using blue-sensitive facilities. Currently this is possible with MODS on LBT, but in the near future the upgraded FORS2 on VLT operating from 3400 \AA \ \citep{Boffin_2020}, will also be an ideal instrument. We plan therefore to validate our conclusion by observing and possibly directly detecting the LyC emission in our galaxies, and we have already been awarded time on the LBT to follow up the UDS sources.
If we indeed confirm that \textrm{C}\textsc{iv} emitters are good ionizers, having both high photon production and high escape fraction, this could have important implications to understand the epoch of reionization, where obviously the LyC flux is not observable directly: indeed Ly$\alpha$, one of the best indirect indicators of LyC escape, looses visibility when the IGM starts to be highly neutral, as it is easily suppressed by neutral hydrogen and is visible only in highly ionized peculiar regions \citep{Castellano_2022}. Nebular \textrm{C}\textsc{iv} emission on the other hand would be easily observable with \textit{JWST/NIRSpec} in star forming galaxies up to very high redshift and could therefore be used as a valid alternative to pinpoint the sources responsible for cosmic reionization. 

\begin{acknowledgements}

The VANDELS Data Release 4 (DR4) is now publicly available and can be accessed using the VANDELS database at \url{http://vandels. inaf.it/dr4.html}, or through the ESO archives. The data published in this paper have been obtained using the \texttt{pandora.ez} software developed by INAF IASF-Milano. The
data analysis work has made extensive use of Python packages \texttt{astropy} \citep{Astropy_2018}, \texttt{numpy} \citep{Harris_2020}, and \texttt{Matplotlib} \citep{4160265}. The codes may be shared upon reasonable written request to the corresponding author. MLl acknowledges support from the ANID/Scholarship Program/Doctorado Nacional/2019-21191036.
\end{acknowledgements}

%
 \bibliographystyle{aa} 
 \bibliography{bibliography.bib} 
%
\newpage
\section{Appendix}

\begin{table}[ht]
\caption{Set of \textsc{BEAGLE} parameters used to fit both the parent and the \textrm{C}\textsc{iv} samples' sources). \label{tab:BEAGLE}}
\centering
\begin{tabular}{llll}
\hline \hline
\textbf{Parameter} & \textbf{type} & \textbf{info} & \textbf{range} \\ 
\hline
\small{sfh\_type} & \small{fixed} & \small{value: delayed}  & \\
\small{tau} & \small{fitted} & & \small{[7.0, 10.5]} \\
\small{mass} & \small{fitted} & & \small{[7.0, 12.0]} \\
\small{max\_stellar\_age} & \small{fitted} & & \small{[7.0, 10.0]} \\
\small{formation\_redshift} & \small{fitted} & & \small{[7.011499, 50.0]}\\
\small{metallicity} & \small{fitted} & & \small{[-2.2, 0.25]} \\
\small{sfr} & \small{fitted} & & \small{[0.0, 3.0]} \\
\small{current\_sfr\_timescale} & \small{fixed} & \small{value: 7.0} & \\
\small{redshift} & \small{fitted}  & & \\
\small{attenuation\_type} & \small{fixed}  & \small{value: CF00} & \\
\small{tauV\_eff}  & \small{fitted}  & & \small{[0.001, 5.01]} \\
\small{mu} & \small{fixed}  & \small{value}: 0.4  & \\
\small{nebular\_logU}  & \small{fitted}  & & \small{[-4.0, -1.0]} \\
\small{nebular\_xi}  & \small{fitted}  & & \small{[0.1, 0.5]}  \\
\small{nebular\_Z}  & \small{dependent}  & & \\
\small{nebular\_sigma}  & \small{fixed}  & \small{value: 22.93} & \\
\hline

\end{tabular}
\end{table}

\end{document}